\documentclass[aps,prd,onecolumn,showpacs,showkeys,amsmath,amssymb]{revtex4}
\usepackage{amsfonts}
\usepackage{amsmath}
\usepackage{graphicx}
\usepackage{subfig}
\usepackage{dcolumn}
\usepackage{natbib}
\usepackage{bm}
\usepackage{booktabs}
\usepackage{slashed}

\usepackage{ulem}
\usepackage{color}

\makeatletter
\newcommand{\figcaption}{\def\@captype{figure}\caption}
\newcommand{\tabcaption}{\def\@captype{table}\caption}

\newcommand{\Rmnum}[1]{\expandafter\@slowromancap\romannumeral #1@}
\def\hlinewd#1{%
  \noalign{\ifnum0=`}\fi\hrule \@height #1 \futurelet
   \reserved@a\@xhline}
\makeatother
\def\dab{\int^{\alpha_{max}}_{\alpha_{min}}d\alpha\int^{\beta_{max}}_{\beta_{min}}d\beta}
\def\qq{\langle\bar qq\rangle}
\def\GGa{\langle GG\rangle}
\def\GGb{\langle \alpha_sGG\rangle}
\def\GGGa{\langle GGG\rangle}
\def\GGGb{\langle g_s^3fGGG\rangle}
\def\JJa{\langle jj\rangle}
\def\JJb{\langle g_s^4jj\rangle}
\def\f(s){[(\alpha+\beta)m_c^2-\alpha\beta s]}
\def\non{\\ \nonumber}

\begin{document}

\title{Mass Spectrum of Heavy Quarkonium Hybrids}

\author{Wei Chen}
\email{wec053@mail.usask.ca}
\author{R. T. Kleiv}
\email{robin.kleiv@usask.ca}
\author{T. G. Steele}
\email{tom.steele@usask.ca}
\affiliation{Department of Physics
and Engineering Physics, University of Saskatchewan, Saskatoon, SK, S7N 5E2, Canada}
\author{B. Bulthuis}
\author{D. Harnett}
\email{derek.harnett@ufv.ca}
\author{J. Ho}
\author{T. Richards}

\affiliation{Department of Physics, University of the Fraser Valley, Abbotsford, BC, V2S 7M8, Canada}
\author{Shi-Lin Zhu} 
\email{zhusl@pku.edu.cn}
\affiliation{Department of Physics
and State Key Laboratory of Nuclear Physics and Technology\\
Peking University, Beijing 100871, China }

\begin{abstract}
We have extended the calculation of the correlation functions of heavy quarkonium hybrid operators with various $J^{PC}$ quantum numbers to include QCD condensates up to dimension six. In contrast to previous analyses which were unable to optimize the QCD sum-rules for certain $J^{PC}$, recent work has shown that inclusion of dimension six condensates stabilizes the hybrid sum-rules and permits reliable mass predictions. In this work we have investigated the effects of the dimension six condensates on the remaining channels. After performing the QCD sum-rule analysis, we update the mass spectra of charmonium and bottomonium hybrids with exotic and non-exotic quantum numbers. We identify that the negative-parity states with $J^{PC}=(0, 1, 2)^{-+}, 1^{--}$ form the lightest hybrid supermultiplet while the positive-parity states with $J^{PC}=(0, 1)^{+-}, (0, 1, 2)^{++}$ belong to a heavier hybrid supermultiplet, confirming the supermultiplet structure found in other approaches. The hybrid with 
$J^{PC}=0^{--}$ has a much higher mass which may suggest a different excitation of the gluonic field compared to other channels. In agreement with previous results, we find that the $J^{PC}=1^{++}$ charmonium hybrid is substantially heavier than the $X(3872)$, which seems to preclude a pure charmonium hybrid interpretation for this state.
\end{abstract}



\maketitle
\section{Introduction}\label{sec:INTRODUCTION}
In the constituent quark model, hadrons are described as $q\bar q$ mesons and $qqq$ baryons. Most of the experimentally observed
resonances can be accommodated in the quark model~\cite{1985-Godfrey-p189-231, 2012-Beringer-p10001-10001}.
However, QCD itself allows a much richer hadron spectrum, such as multiquarks, hybrids, glueballs, {\it etc...}~\cite{2007-Klempt-p1-202}.
Hybrid mesons ($\bar qgq$) are composed of a color-octet quark-antiquark pair and an excited gluonic field. Since the excited gluonic field could
carry quantum numbers other than $0^{++}$, hybrids can appear with both the ordinary (the same as conventional $q\bar q$ mesons)
and exotic quantum numbers ($J^{PC}=0^{--}, 0^{+-}, 1^{-+}, 2^{+-},\ldots$). The states with the exotic quantum numbers are not accessible
for a $q\bar q$ state. Hybrids provide a good platform to search for these exotic quantum numbers.

The light hybrids were studied in the MIT bag model~\cite{1983-Barnes-p241-241, 1983-Chanowitz-p211-211}, in which the hybrids with
$J^{PC}=(0, 1, 2)^{-+}, 1^{--}$ were predicted to form the lightest hybrid supermultiplet consisting of a S-wave color-octet quark-antiquark
pair coupled to an excited gluonic field with $J_g^{P_gC_g}=1^{+-}$. A higher hybrid supermultiplet composed of a P-wave $q\bar q$ pair and the same gluonic excitation would contain states with
$J^{PC}=0^{+-}, (1^{+-})^3, (2^{+-})^2, 3^{+-}, (0, 1, 2)^{++}$, where the superscript denotes the number of such states~\cite{2012-Liu-p126-126, 2011-Dudek-p74023-74023}.
Many other methods such as the flux tube model~\cite{1985-Isgur-p869-869, 1995-Close-p1706-1709, 1995-Barnes-p5242-5256, 1999-Page-p34016-34016, 1995-Close-p233-254}, lattice QCD~\cite{1999-McNeile-p264-266, 1999-Lacock-p261-263, 2003-Bernard-p74505-74505, 2005-Hedditch-p114507-114507} and QCD sum rules~\cite{1983-Govaerts-p262-262, 1984-Govaerts-p1-1, 1984-Latorre-p169-169, 1987-Latorre-p347-347, 1986-Balitsky-p265-273, 2003-Jin-p14025-14025, 2000-Chetyrkin-p145-150, 1999-Zhu-p97502-97502} were also used to study the light hybrids. To date, there is some evidence of the exotic light hybrid with $J^{PC}=1^{-+}$~\cite{1997-Thompson-p1630-1633, 1998-Abele-p175-184, 1999-Abele-p349-355, 2007-Adams-p27-31}.

For heavy quarkonium hybrids, which hereafter we shall refer to as heavy quark hybrids, calculations including the constituent gluon model~\cite{1978-Horn-p898-898}, the flux tube
model~\cite{1995-Barnes-p5242-5256}, QCD sum rules~\cite{1985-Govaerts-p215-215,1985-Govaerts-p575-575,1987-Govaerts-p674-674,
1999-Zhu-p31501-31501,2012-Qiao-p15005-15005, 2012-Harnett-p125003-125003, 2012-Berg-p34002-34002} and lattice QCD~\cite{1990-Perantonis-p854-868, 1999-Juge-p4400-4403, 2006-Liu-p54510-54510, 2006-Luo-p34502-34502, 2011-Liu-p140-140, 2012-Liu-p126-126} have been performed. The hybrid supermultiplet structures described above were confirmed for the heavy quark sector in lattice QCD~\cite{2012-Liu-p126-126} and the P-wave quasigluon approach~\cite{2008-Guo-p56003-56003}. So far, however, no definitive experimental signal for heavy hybrid mesons has been observed, although many unexpected charmonium-like and bottomonium-like states have been discovered in the past several years~\cite{2011-Brambilla-p1534-1534, 2010-Nielsen-p41-83, 2006-Swanson-p243-305,
2008-Zhu-p283-322, 2007-Rosner-p12002-12002}. Some of these states, called $X,\, Y,\, Z$, do not fit in the conventional quark model easily
and are considered to be candidates for exotic states beyond the quark model, such as molecular states, tetraquarks, baryonium
and quarkonium hybrids. For example, $Y(4260)$ was interpreted as a charmonium hybrid state in
Refs.~\cite{2005-Zhu-p212-212, 2005-Close-p215-222, 2007-Klempt-p1-202}. In Ref.~\cite{2003-Close-p210-216}, $X(3872)$ was also
proposed to be a charmonium hybrid. To search for hybrids in the heavy quarkonium region further theoretical investigations of the hybrid
spectrum are still needed.

QCD sum rules is a very powerful non-perturbative method~\cite{1979-Shifman-p385-447, 1985-Reinders-p1-1, 2000-Colangelo-p1495-1576} which has been widely used to study hadron structures. In Refs.~\cite{1985-Govaerts-p215-215,1985-Govaerts-p575-575,1987-Govaerts-p674-674}, Govaerts \textit{et al.} performed a QCD sum-rule analysis of the heavy quark hybrids including leading-order contributions up to the dimension four gluonic condensate.
The sum rules of some $J^{PC}$ channels were unstable and thus these mass predictions were unreliable. Recently, the $J^{PC}=1^{--}$~\cite{2012-Qiao-p15005-15005}, $1^{++}$~\cite{2012-Harnett-p125003-125003} and $0^{-+}$~\cite{2012-Berg-p34002-34002} channels have been re-analyzed by including the dimension six tri-gluon condensate. The dimension six contributions have proven to be very important because they stabilize the hybrid sum rules and allow mass prediction to be made. In this work, we will study the mass spectrum of heavy quarkonium hybrids using the QCD sum rule method. We will calculate the dimension six condensates including $\GGGb$ and the light-quark condensate  $\JJb$, where the latter condensate has not previously been calculated in the hybrid sum rules.

The paper is organized as follows. In Sec.~\Rmnum{2}, we calculate the correlation functions and spectral densities of the hybrid operators with various quantum numbers and collect them in Appendix~\ref{sec:rhos}. In Sec.~\Rmnum{3}, we perform the numerical analysis and extract masses of charmonium and bottomonium hybrids. In the last section we summarize our results and comment on their implications for heavy quarkonium spectroscopy.

\section{Laplace Sum Rules for the Heavy Quarkonium Hybrids}\label{sec:QSR}
We consider the two-point correlation function
\begin{eqnarray}
\Pi_{\mu\nu}(q)= i\int
d^4x \,e^{iq \cdot x}\,\langle0|T[J_{\mu}(x)J_{\nu}^{\dag}(0)]|0\rangle, \label{equ:Pi}
\end{eqnarray}
where $J_{\mu}$ is the interpolating current. In this work, we use the following hybrid interpolating currents coupling to various quantum numbers:
\begin{eqnarray}
\nonumber
J_{\mu}&=&g_s\bar Q\frac{\lambda^a}{2}\gamma^{\nu}G^a_{\mu\nu}Q,~~~~~~~J^{PC}=1^{-+}, 0^{++},
\\
J_{\mu}&=&g_s\bar Q\frac{\lambda^a}{2}\gamma^{\nu}\gamma_5G^a_{\mu\nu}Q,~~~~J^{PC}=1^{+-}, 0^{--}, \label{currents}
\non
J_{\mu\nu}&=&g_s\bar Q\frac{\lambda^a}{2}\sigma_{\mu}^{\alpha}\gamma_5 G^a_{\alpha\nu}Q,~~~~J^{PC}=2^{-+}, 1^{++}, 1^{-+}, 0^{-+}\,,
\end{eqnarray}
in which $Q$ represents a heavy quark ($c$ or $b$), $g_s$ is the strong coupling, $\lambda^a$ are the Gell-Mann matrices and $G^a_{\mu\nu}$ is the gluon field strength. 
By replacing $G^a_{\mu\nu}$ with $\tilde G^a_{\mu\nu}=\frac{1}{2}\epsilon_{\mu\nu\alpha\beta}G^{\alpha\beta,a}$,
we can also obtain the corresponding operators with opposite parity.
These hybrid correlation functions were originally studied in Refs.~\cite{1985-Govaerts-p215-215,1985-Govaerts-p575-575,1987-Govaerts-p674-674} in which the perturbative and gluon condensate contributions have been calculated to perform the QCD sum rule analysis. According to these calculations, the stability of the sum rules depends on the relative sign of the gluon condensate and the perturbative term. Only a  relative negative sign results in stable sum rules. As shown in Refs.~\cite{1985-Govaerts-p215-215,1985-Govaerts-p575-575,1987-Govaerts-p674-674}, the sum rules for $J^{PC}=0^{--}, 0^{++}, 1^{+-}, 1^{++}, 2^{++}$ channels are stable while the  $J^{PC}=0^{-+}, 0^{+-}, 1^{-+}, 1^{--}, 2^{-+}$ channels are unstable. In this work, we will calculate the dimension six condensates in the correlation functions and show that
they stabilize the hybrid sum rules for all unstable $J^{PC}=0^{-+}, 0^{+-}, 1^{-+}, 1^{--}, 2^{-+}$ channels, confirming the stabilizing effect of dimension six condensates previously found for $1^{--}$ \cite{2012-Qiao-p15005-15005}, $1^{++}$~\cite{2012-Harnett-p125003-125003} and $0^{-+}$~\cite{2012-Berg-p34002-34002}.

For the hybrid operators in Eq.~(\ref{currents}), the two-point correlation functions have the following structures:
\begin{eqnarray}
i\int d^4x \,e^{iq\cdot x}\,\langle 0|T [J_\mu(x)J_\nu^\dagger(0)]|0\rangle
&=&\left[\frac{q_\mu q_\nu}{q^2}-g_{\mu\nu}\right]\Pi_V(q^2)+\frac{q_\mu q_\nu}{q^2}\Pi_S(q^2), \label{pivector}\\
i\int d^4x\,e^{iq\cdot x}\,\langle0|T[J_{\mu\nu}(x)J_{\rho\sigma}^{\dag}(0)]|0\rangle
&=&\left[\eta_{\mu\rho}\eta_{\nu\sigma}+\eta_{\mu\sigma}\eta_{\nu\rho}
-\frac{2}{3}\eta_{\mu\nu}\eta_{\rho\sigma}\right]\Pi_T(q^2)+\ldots \,, \label{pitensor}
\end{eqnarray}
where $\eta_{\mu\nu}=q_{\mu}q_{\nu}/q^2-g_{\mu\nu}$. $\Pi_V(q^2)$, $\Pi_S(q^2)$ in Eq.~(\ref{pivector}) and $\Pi_T(q^2)$ in Eq.~(\ref{pitensor}) are the invariant structures referring to pure spin-1, spin-0 and spin-2 states, respectively. In Eq.~(\ref{pitensor}), the invariant structures for spin-0 and spin-1 are not written out explicitly because
we will not consider contributions arising from these terms in this paper.

At the hadron level, the correlation function can be described by the dispersion relation,
\begin{eqnarray}
\Pi(q^2)=(q^2)^N\int_{4m^2}^{\infty}\frac{\rho(s)}{s^N(s-q^2-i\epsilon)}ds+\sum_{n=0}^{N-1}b_n(q^2)^n.
\end{eqnarray}
The summation on the right hand side represents the subtraction terms which can be removed by taking the Borel transform of $\Pi(q^2)$. The spectral function $\rho(s)$ is defined using the pole plus continuum approximation
\begin{eqnarray}
\rho(s)\equiv\sum_n\delta(s-m_n^2)\langle0|J_{\mu}|n\rangle\langle n|J_{\mu}^{\dagger}|0\rangle
=f_X^2m_X^8\delta(s-m_X^2)+ \mbox{continuum},  \label{Phenrho}
\end{eqnarray}
where the intermediate states $n$ must have the same quantum numbers as the interpolating currents $J_{\mu}$,
$m_X$ denotes the mass of the lowest lying resonance and the dimensionless quantity $f_X$ is the coupling of the resonance to the current
\begin{eqnarray}
\langle0|J_{\mu}|X\rangle&=&f_Xm_X^4\epsilon_{\mu}\,, \\
\langle0|J_{\mu\nu}|X\rangle&=&f_Xm_X^4\epsilon_{\mu\nu}\,. \;
\end{eqnarray}

\begin{figure}
\includegraphics[scale=0.6]{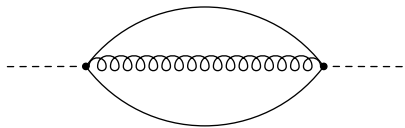}
\figcaption{Feynman diagram representing the perturbative contribution to the correlation functions. Solid and curly lines represent quark and gluon propagators  respectively, while the dashed line represents the interpolating current.} \label{fig1}
\end{figure}

\begin{figure}
\includegraphics[scale=0.6]{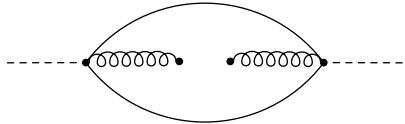}
\figcaption{Feynman diagram representing the $\langle\alpha_sGG\rangle$
contribution to the correlation functions.} \label{fig2}
\end{figure}

\begin{figure}
\includegraphics[scale=0.6]{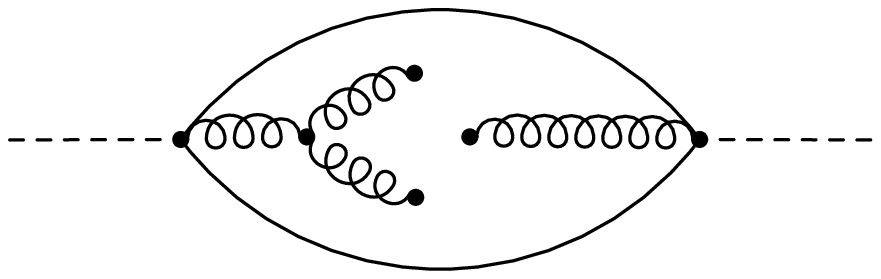}
 \includegraphics[scale=0.6]{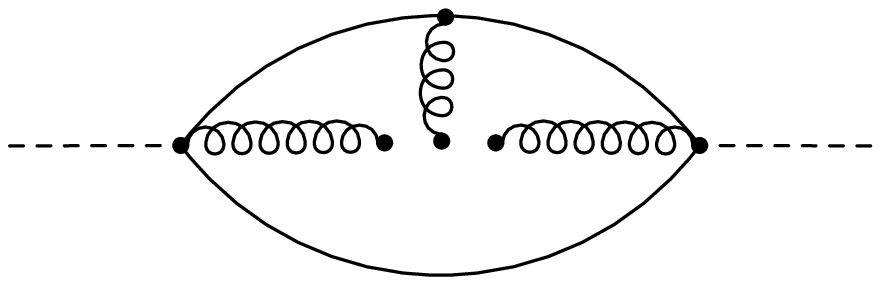}
 \centerline{\hspace{-0.04in} {(a)} \hspace{1.90in}{ (b)}}
 \figcaption{Feynman diagrams representing the $\langle DDG\rangle$ (a) and $\GGGb$ (b)
contributions to the correlation functions.
} \label{fig3}
\end{figure}

At the quark-gluon level, the correlation function can be computed via the operator product expansion (OPE). For heavy quark hybrids, the quark condensates are expressed in terms of the gluon condensate via the heavy quark mass expansion and hence give no contributions to the correlation function. To calculate the correlation function, only the perturbative diagram (Fig.~\ref{fig1}), gluon condensate diagram (Fig.~\ref{fig2}) and diagrams for dimension six condensates (Fig.~\ref{fig3}) are involved. In Fig~\ref{fig3}, the dimension six condensates contain $\GGGb$ (Fig.~\ref{fig3}b) and $\langle DDG\rangle$ (Fig.~\ref{fig3}a). Using the equation of motion, Bianchi identities and commutation relations, the condensate $\langle DDG\rangle$ can be expressed in terms of $\GGGb$ and $\JJb$. 

We have calculated the Wilson coefficients using two independent techniques, finding complete agreement between them. 
First, we have used the same approach that was utilized in Refs.~\cite{2012-Harnett-p125003-125003,2012-Berg-p34002-34002} for $0^{-+}$ and $1^{++}$ heavy hybrids, respectively. In this approach, we utilize the Mathematica package Tarcer~\cite{1998-Mertig-p265-273}, which implements the $d$ dimensional recurrence relations developed in Refs.~\cite{1996-Tarasov-p6479-6490,1997-Tarasov-p455-482} to reduce 
the number of distinct integrals that must be calculated. The resulting minimal set of loop integrals can be evaluated using results given in 
Refs.~\cite{1991-Boos-p1052-1063,1992-Davydychev-p358-369,1993-Broadhurst-p287-302}, leading to very compact expressions for the Wilson coefficients in terms of generalized hypergeometric functions. Second, we have utilized the technique used in Refs.~\cite{2010-Chen-p105018-105018, 2011-Chen-p34010-34010, 2011-Chen-p-,2013-Du-p14003-14003}. With this method the perturbative spectral densities are expressed as integrations that can be performed numerically. In this approach we use the momentum space quark propagator
\begin{eqnarray}
iS_{ab}(p) &=&
\frac{i\delta_{ab}}{\slashed{p}-m}+\frac{i}{4}g_s\frac{\lambda^n_{ab}}{2}G_{\mu\nu}^n
\frac{\sigma^{\mu\nu}(\slashed{p}+m)+(\slashed{p}+m)\sigma^{\mu\nu}}{(p^2-m^2)^2},
\end{eqnarray}
where $\sigma^{\mu\nu}=\frac{i}{2}\left[\gamma^\mu\,,\gamma^\nu\right]$, and $a\,,b$ are color indices. To our knowledge, this is the first time that these two calculational approaches have been directly compared, therefore the agreement between them is noteworthy. In addition, we interpret this agreement as a very robust test of the veracity of our results.


The correlation function obtained at the quark-gluon level should be equivalent to that described at the hadron level due to quark-hadron duality, establishing a
sum rule to extract the hadron mass. The Borel transform is applied to the correlation functions at both levels to pick out the lowest lying resonance, eliminate dispersion-relation subtractions, and enhance the OPE convergence. Using the spectral function in Eq.~(\ref{Phenrho}), we arrive at
\begin{eqnarray}
\mathcal{L}_{k}\left(s_0, M_B^2\right)=f_X^2 m_X^{8+2k}e^{-m_X^2/M_B^2}=\int_{4m^2}^{s_0}ds\,s^k\,\rho(s)\,e^{-s/M_B^2}\,,
\label{sumrule}
\end{eqnarray}
where $s_0$ is the continuum threshold parameter and $M_B$ is the Borel mass. Then the hadron mass can be extracted using 
\begin{eqnarray}
m_X^2=\frac{\mathcal{L}_{1}\left(s_0\,, M_B^2\right)}{\mathcal{L}_{0}\left(s_0\,, M_B^2\right)}\,,
\label{mass}
\end{eqnarray}
in which $m_X$ denotes the heavy hybrid mass.

For all the interpolating currents in Eq.~(\ref{currents}), we calculate the correlation functions and the spectral densities are then
obtained via $\rho(s)=\frac{1}{\pi}$Im$\Pi(s)$. We list the expressions for the correlation functions and the spectral densities in Appendix~\ref{sec:rhos}. As mentioned above, we have used two distinct methods to determine these expressions for the spectral functions. First, we have used the same method as in Refs.~\cite{2012-Harnett-p125003-125003,2012-Berg-p34002-34002} to calculate the full correlation function, from which the imaginary part can be extracted through analytic continuation and hence a closed form expression for the spectral function can be constructed. Second, we have used the method of Refs.~\cite{2010-Chen-p105018-105018, 2011-Chen-p34010-34010, 2011-Chen-p-, 2013-Du-p14003-14003} which can be used to determine integral representations of the spectral functions. We have numerically verified that these two approaches lead to identical results. This provides a very strong test of our results.  Another check is that our results for the perturbative 
and gluon condensate contributions are in numerical agreement with those in Refs.~\cite{1985-Govaerts-p215-215,1985-Govaerts-p575-575,1987-Govaerts-p674-674}.

Besides the perturbative term and the dimension four gluon condensate $\GGb$, we also calculate the dimension six condensates $\GGGb$ and $\JJb$. The condensate $\JJb$ has not previously been evaluated in the hybrid sum rules. As mentioned above, it comes from applying  equations of motion in the condensate $\langle DDG\rangle$ arising from the diagram Fig.~(\ref{fig3}a). Within the vacuum factorization assumption, $\JJb$ is proportional to the square of $g_s^2\qq$. Although the numerical analysis shows that $\JJb$ is small enough to be ignored compared to $\GGGb$, we include its contribution in our analysis because of the important stabilizing role of the dimension six condensates. We have only included the integral representations for these contributions in Appendix~\ref{sec:rhos}.

From the formulae for the spectral densities in Appendix~\ref{sec:rhos}, we find that the perturbative contributions are invariant and the gluon condensate and tri-gluon condensate contributions change sign under parity reversal. This is in agreement with the results for the perturbative 
and gluon condensate contributions in Refs.~\cite{1985-Govaerts-p215-215,1985-Govaerts-p575-575,1987-Govaerts-p674-674}. It should be noted 
that both the expression for $\rho^{\GGGa}$ and that of $\rho^{\JJa}$ have a singularity at $s=4m^2$.
One should be very cautious in calculating the Borel transforms of the correlation functions from these expressions due to these singularities. In Refs.~\cite{2012-Harnett-p125003-125003, 2012-Berg-p34002-34002} a numerical limiting procedure was developed based on the full analytic structure of the correlation functions and the relation between the Borel transform and inverse Laplace transform. This limiting procedure serves to cancel the integration divergences at $s=4m^2$ in the $\eta\to 0$ limit. A suitable value of $\eta$ should be determined to perform the numerical analysis. Alternatively, these divergences can also be eliminated if we use the integral forms of $\rho^{\GGGa}$ and $\rho^{\JJa}$ in Eqs.~(\ref{SP1-+})--(\ref{last_eqn}) where the delta functions absorb divergences of the spectral densities at $x=0$ and $x=1$. The $\eta\to 0$ limiting expressions are necessary when the closed form expressions for the spectral densities are used (see Appendix~\ref{sec:rhos}).

\section{Numerical Analysis}\label{sec:NA}
To perform the QCD sum rule numerical analysis, we use the following values of the heavy quark masses
and the condensates~\cite{2009-Chetyrkin-p74010-74010, 2012-Narison-p259-263, 2010-Narison-p559-559,
2007-Kuhn-p192-215}: 
\begin{eqnarray}
\nonumber &&m_c(\mu=m_c)=\overline m_c=(1.28\pm 0.02)~\mbox{GeV},
\non &&m_b(\mu=m_b)=\overline m_b=(4.17\pm 0.02)~\mbox{GeV},
\non &&\GGb=(7.5\pm2.0)\times 10^{-2}~\mbox{GeV}^4,
\\&& \GGGb=-(8.2\pm1.0)~\mbox{GeV}^2\GGb,
\non &&\qq=-(0.23\pm0.03)^3~\mbox{GeV}^3,
\non &&\JJb=-\frac{4}{3}g_s^4\qq^2,
\label{inputs}
\end{eqnarray}
in which the charm and bottom quark masses are the running masses in the $\overline{\rm MS}$ scheme.
Note that there is a minus sign implicitly included in the definition of the coupling constant $g_s$ in this work.
We choose the renormalization scale $\mu=\overline m_c$ for the charmonium systems and $\mu=\overline m_b$ for bottomonium systems, but other choices of renormalization scale for the $\overline{\rm MS}$ masses can be expressed via the leading order expressions:
\begin{eqnarray}
m_c(\mu)=\overline m_c\bigg(\frac{\alpha_s(\mu)}{\alpha_s(\overline m_c)}\bigg)^{12/25},
\\m_b(\mu)=\overline m_b\bigg(\frac{\alpha_s(\mu)}{\alpha_s(\overline m_b)}\bigg)^{12/23}\,.
\end{eqnarray}
For the hybrid charmonium and bottomonium analyses, the strong coupling is then determined by evolution
from the $\tau$ and $Z$ masses, respectively:
\begin{eqnarray}
\alpha_s(\mu)&=&\frac{\alpha_s(M_{\tau})}{1+\frac{25\alpha_s(M_{\tau})}{12\pi}\log(\frac{\mu^2}{M_{\tau}^2})}, \quad \alpha_s(M_{\tau})=0.33; \label{alpha_cc}
\\ \alpha_s(\mu)&=&\frac{\alpha_s(M_{Z})}{1+\frac{23\alpha_s(M_{Z})}{12\pi}\log(\frac{\mu^2}{M_{Z}^2})}, \quad \alpha_s(M_{Z})=0.118,
\label{alpha_bb}
\end{eqnarray}
in which the $\tau$ and $Z$ masses, $\alpha_s(M_{\tau})$ and $\alpha_s(M_{Z})$ are from the Particle Data Group~\cite{2012-Beringer-p10001-10001}.
Apart from small chiral-violating effects, the QCD condensate $\GGb$ is renormalization scale invariant.  Although $\JJb$ has complicated renormalization-group behaviour \cite{1983-Narison-p217-217}, these effects are negligible because the contribution from this condensate is a small numerical effect.  Finally, the condensate $ \GGGb$ in \eqref{inputs} is determined at the scale $\mu=\overline m_c$ \cite{2012-Narison-p259-263} so our charmonium analysis is self-consistent and does not require renormalization-group evolution of  $ \GGGb$.  For bottomonium systems,  the effect of the $ \GGGb$ anomalous dimension  \cite{1983-Narison-p217-217} has been found to be  negligible  \cite{2002-Narison-p1-1}.

We begin with the sum-rule analysis of the hybrid charmonium mass spectrum. The stability of the QCD sum rule requires suitable working regions
for the continuum threshold parameter $s_0$ and Borel mass $M_B$. According to Eq.~(\ref{sumrule}), excited state contributions are naturally
suppressed by the exponential weight function for small $M_B^2$. The lowest lying resonance contribution will be enhanced in the same region.
On the other hand, however, the OPE convergence suffers if $M_B^2$ is too small. In our analysis, the Borel mass working region is determined by the convergence of the OPE series and the pole contribution. The requirement of OPE convergence determines the lower bound on $M_B^2$ while the pole contribution constraint leads to the upper bound.

For hybrid charmonium the dominant nonperturbative contributions come from the gluon condensate $\GGb$.
However, the dimension six condensates are also important because they can stabilize the mass sum rules~\cite{2012-Qiao-p15005-15005,2012-Berg-p34002-34002,2012-Harnett-p125003-125003}. We have
calculated dimension six condensates $\GGGb$ and $\JJb$ in the spectral densities, extending the results of \cite{1985-Govaerts-p215-215,1985-Govaerts-p575-575,1987-Govaerts-p674-674}. From the numerical analysis, we see that the
condensate $\JJb$ is much smaller than $\GGGb$ (Fig.~\ref{figOPEcc1-+}). To study the OPE convergence, we require
that the gluon condensate be less than one third of the perturbative term while the tri-gluon condensate be less than
one third of the gluon condensate. This requirement gives the lower bound on $M_B^2$. For example, we show the OPE
convergence for the exotic channel $J^{PC}=1^{-+}$ in Fig.~\ref{figOPEcc1-+}, from which we obtain the lower bound
of the Borel mass $M_{Bmin}^2=4.6$ GeV$^2$.
\begin{center}
\begin{tabular}{c}
\scalebox{0.8}{\includegraphics{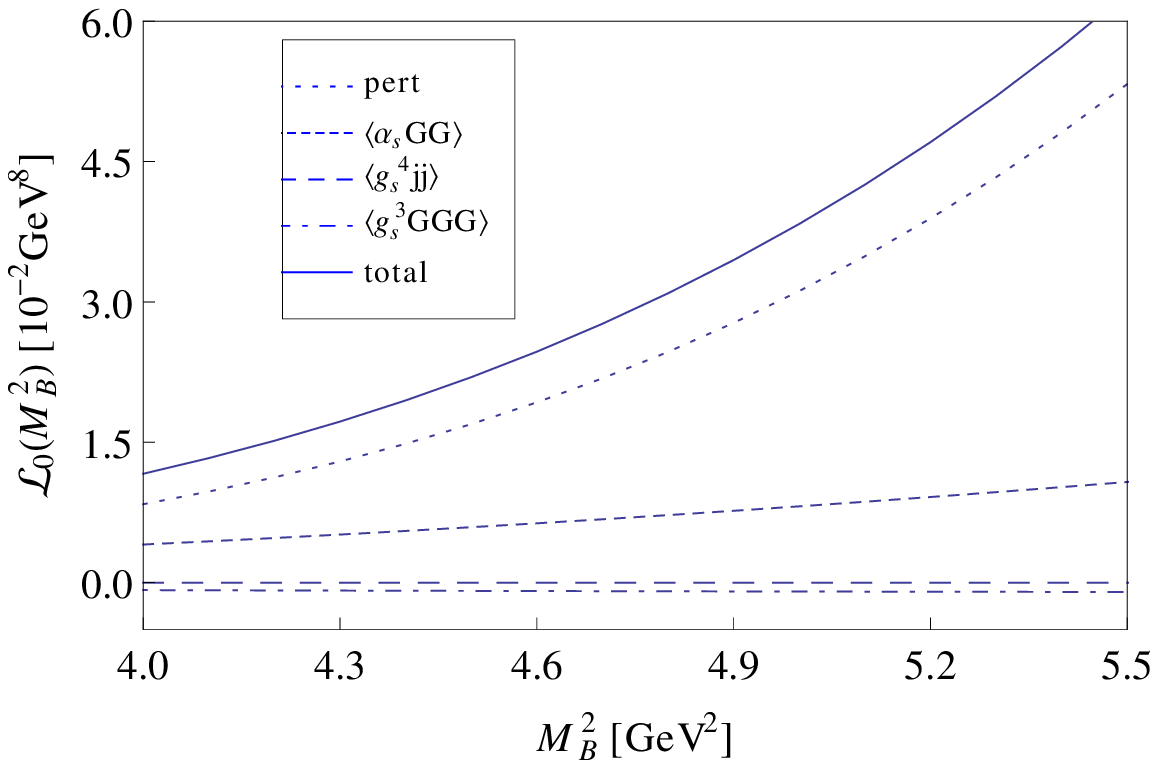}}
\end{tabular}
\figcaption{The contributions of each term in the OPE series, including the perturbative term, gluon condensate $\GGb$, tri-gluon condensate $\GGGb$ and
the condensate $\JJb$ for the $1^{-+}$ charmonium hybrid sum rule ${\cal L}_0(M_B^2)$ with $s_0\to\infty$.} \label{figOPEcc1-+}
\end{center}

We define the pole contribution (PC) as:
\begin{eqnarray}
\mbox{PC}(s_0, M_B^2)=\frac{\int_{4m^2}^{s_0}\rho(s)e^{-s/M_B^2}ds}{\int_{4m^2}^{\infty}\rho(s)e^{-s/M_B^2}ds}, \label{PC}
\end{eqnarray}
where PC is a function of $s_0$ and $M_B^2$. In order to study its dependence on $M_B^2$, one should first fix the value
of $s_0$. In order to do so we study the variation of the hybrid mass $m_X$ versus $s_0$, as shown in Fig.~\ref{fig1-+cc} for the $J^{PC}=1^{-+}$
channel. We find that the hybrid mass is very stable for different values of $M_B^2$ around $s_0=17$ GeV$^2$. Using this value
of $s_0$, we obtain the upper bound on the Borel mass of $M_{\rm Bmax}^2=6.5$ GeV$^2$ by requiring the pole contribution be larger
than $10\%$ in Eq.~(\ref{PC}). In Fig.~\ref{fig1-+cc}, we plot the Borel curve of $m_X$ versus $M_B^2$ for the $1^{-+}$ 
charmonium hybrid, which shows that the mass sum rule is very stable in the $M_B^2$ working region. Finally, the predicted
mass of the exotic $J^{PC}=1^{-+}$ hybrid charmonium is $3.70$ GeV. This value is about $0.5$ GeV lower than the lattice
result in Ref.~\cite{2012-Liu-p126-126}.
\begin{center}
\begin{tabular}{lr}
\scalebox{0.6}{\includegraphics{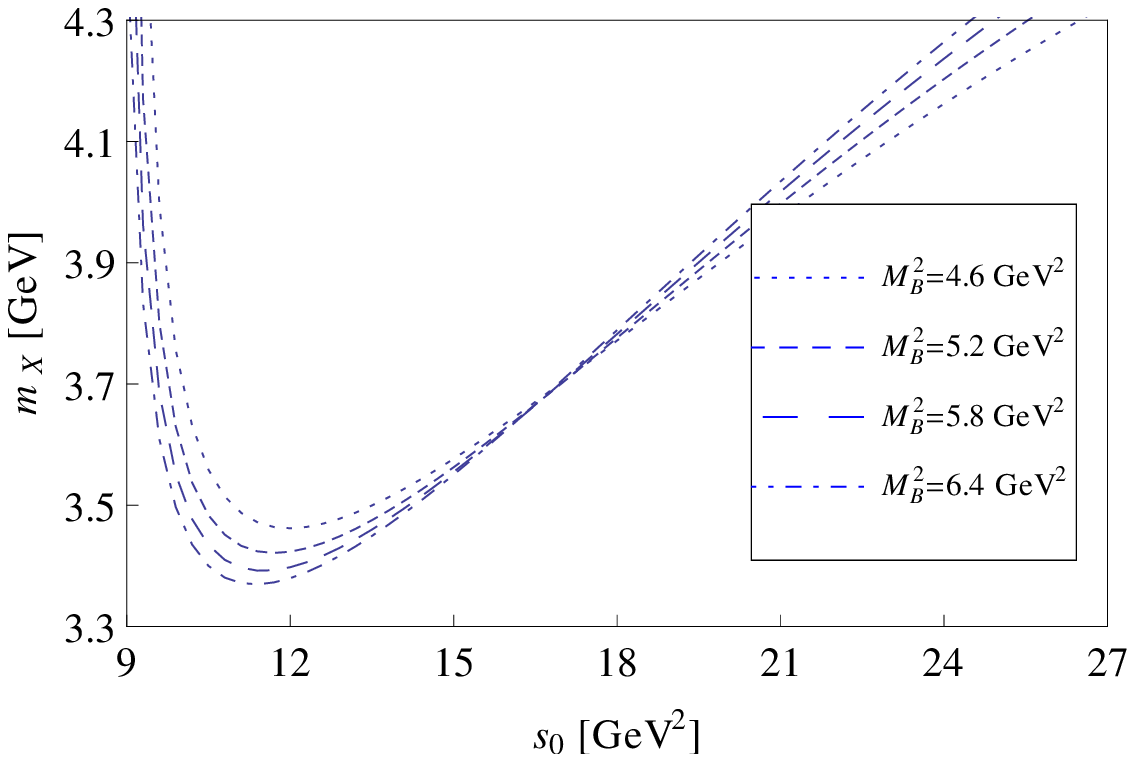}}&
\scalebox{0.6}{\includegraphics{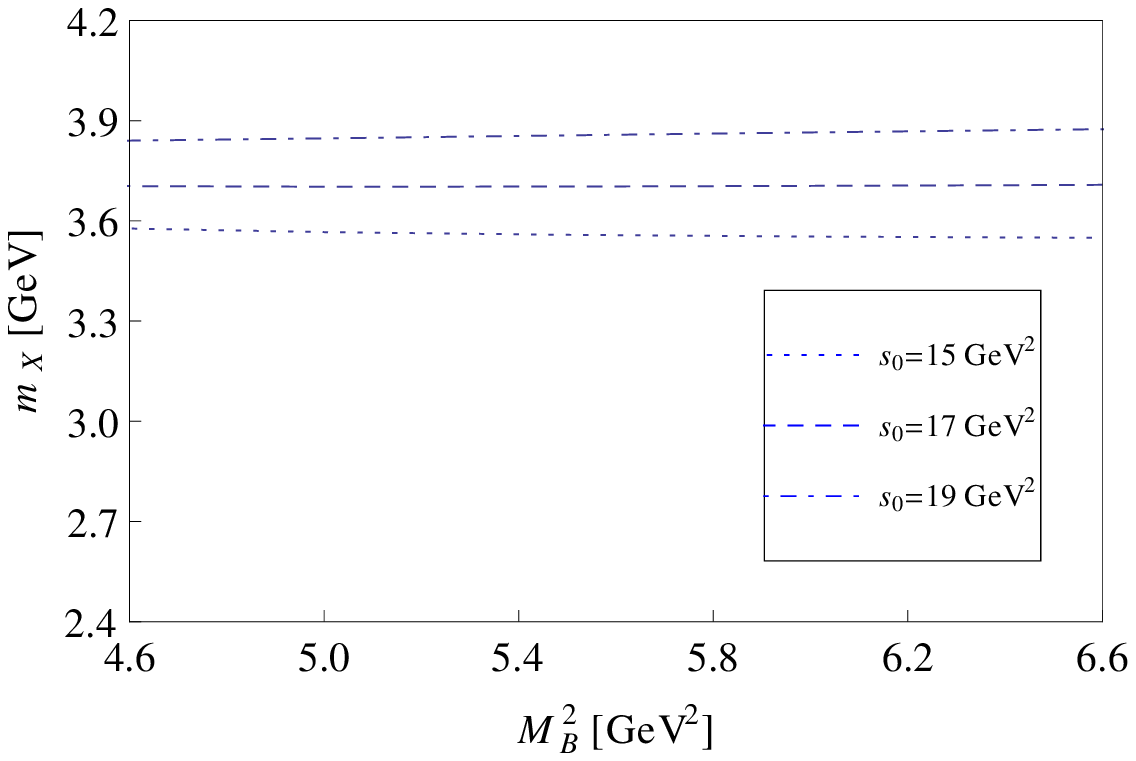}}
\end{tabular}
\figcaption{The variations of $m_X$ with $s_0$ and $M_B^2$ for the $J^{PC}=1^{-+}$ charmonium hybrid.} \label{fig1-+cc}
\end{center}

After performing the QCD sum rule analyses, we extract masses of the charmonium hybrids with various quantum numbers as summarized in Table~\ref{table1}. We also give the corresponding Borel windows, threshold values and pole contributions. Only errors from the uncertainties in the charm quark mass and the condensates are taken into account. We do not consider other possible error sources such as truncation of the OPE series, the uncertainty of the threshold value $s_0$ and the variation of Borel mass $M_B$. In fact, the dominant error is from the uncertainty of the gluon condensate for the channels with $J^{PC}=0^{--}, 0^{++}, 1^{++}, 1^{+-}, 2^{++}$ while the tri-gluon condensate dominates for those with $J^{PC}=0^{-+}, 0^{+-}, 1^{-+}, 1^{--}, 2^{-+}$. The errors from the charm quark mass $m_c$ and $\JJb$ are much smaller.
\begin{center}
\begin{tabular}{cccccc}
\hlinewd{.8pt}
& $J^{PC}$ & $s_0(\mbox{GeV}^2)$&$[M^2_{\mbox{min}}$,$M^2_{\mbox{max}}](\mbox{GeV}^2)$&$m_X$\mbox{(GeV)}&PC(\%)\\
\hline
& $1^{--}$   & 15  & $2.5\sim4.8 $& $3.36\pm0.15$& 18.3\\
& $0^{-+}$   & 16  & $5.6\sim7.0 $& $3.61\pm0.21$& 15.4\\
& $1^{-+}$   & 17  & $4.6\sim6.5 $& $3.70\pm0.21$& 18.8\\
& $2^{-+}$   & 18  & $3.9\sim7.2 $& $4.04\pm0.23$& 26.0
\vspace{5pt}\\
& $0^{+-}$   & 20  & $6.0\sim7.4 $& $4.09\pm0.23$& 15.5\\
& $2^{++}$   & 23  & $3.9\sim7.5$ & $4.45\pm0.27$& 21.5\\
& $1^{+-}$   & 24  & $2.5\sim8.4 $& $4.53\pm0.23$& 33.2\\
& $1^{++}$   & 30  & $4.6\sim11.4$& $5.06\pm0.44$& 30.4\\
& $0^{++}$   & 34  & $5.6\sim14.6$& $5.34\pm0.45$& 36.3
\vspace{5pt}\\
& $0^{--}$    & 35  & $6.0\sim12.3$& $5.51\pm0.50$& 31.0\\
\hline
\hlinewd{.8pt}
\end{tabular}
\tabcaption{Masses of the charmonium hybrid states and the corresponding $s_0$,
Borel windows and pole contributions.
\label{table1}}
\end{center}

\begin{center}
\begin{tabular}{cccccc}
\hlinewd{.8pt}
& $J^{PC}$ & $s_0(\mbox{GeV}^2)$&$[M^2_{\mbox{min}}$,$M^2_{\mbox{max}}](\mbox{GeV}^2)$&$m_X$\mbox{(GeV)}&PC(\%)\\
\hline
& $1^{--}$    & 105  & $11\sim17 $& $9.70\pm0.12$&  17.2\\
& $0^{-+}$   & 104  & $14\sim16 $& $9.68\pm0.29$& 17.3\\
& $1^{-+}$   & 107  & $13\sim19 $& $9.79\pm0.22$& 20.4\\
& $2^{-+}$   & 105  & $12\sim19$& $9.93\pm0.21$& 21.7
\vspace{5pt}\\
& $0^{+-}$   & 114  & $14\sim19 $& $10.17\pm0.22$& 17.6\\
& $2^{++}$   & 120  & $12\sim20$ & $10.64\pm0.33$& 19.7\\
& $1^{+-}$   & 123  & $10\sim21 $& $10.70\pm0.53$& 28.5\\
& $1^{++}$  & 134  & $13\sim27$& $11.09\pm0.60$& 27.7\\
& $0^{++}$  & 137  & $13\sim31$& $11.20\pm0.48$& 30.0
\vspace{5pt}\\
& $0^{--}$    & 142  & $14\sim25$& $11.48\pm0.75$& 24.1\\
\hline
\hlinewd{.8pt}
\end{tabular}
\tabcaption{Masses of the bottomonium hybrid states and the corresponding $s_0$,
Borel windows and pole contributions.\label{table2}}
\end{center}

The results show that for the unstable channels with $J^{PC}=0^{-+}, 0^{+-}, 1^{-+}, 1^{--}, 2^{-+}$ in
Refs.~\cite{1985-Govaerts-p215-215,1985-Govaerts-p575-575,1987-Govaerts-p674-674}, the dimension six condensate $\GGGb$
in the spectral functions can stabilize the systems and make it possible to extract the hybrid masses.
This universal trend confirms the stabilizing effect of dimension six condensates previously observed for the $J^{PC}=1^{--}$~\cite{2012-Qiao-p15005-15005}, $1^{++}$~\cite{2012-Harnett-p125003-125003} and $0^{-+}$~\cite{2012-Berg-p34002-34002} channels.
For these previously unstable channels, the pole contributions are very small even though we consider the dimension six condensates contributions.
The small pole contributions result in the narrow Borel windows compared to the stable channels in
Refs.~\cite{1985-Govaerts-p215-215,1985-Govaerts-p575-575,1987-Govaerts-p674-674} with $J^{PC}=0^{--}, 0^{++}, 1^{++}, 1^{+-}, 2^{++}$.
The $1^{++}$ and $0^{-+}$ results agree with the previous sum-rule analyses including the dimension six gluon condensates
\cite{2012-Berg-p34002-34002,2012-Harnett-p125003-125003}. However, the $1^{--}$ mass is somewhat smaller than in
Ref.~\cite{2012-Qiao-p15005-15005} a result we attribute to the absence of the QCD continuum from the condensates in Ref.~\cite{2012-Qiao-p15005-15005}.
We obtain four charmonium hybrid states with $J^{PC}=(0, 1, 2)^{-+}, 1^{--}$ in the range $3.4\sim3.9$ GeV which are much lower than
the other channels in Table~\ref{table1}. However, the $1^{--}$ channel was examined in Ref.~\cite{2008-Kisslinger-p982341-982341}, finding a mass of 3.66 GeV, which is comparable to our prediction for this channel. In the MIT bag model~\cite{1983-Barnes-p241-241, 1983-Chanowitz-p211-211}, the light
hybrids with these quantum numbers have been predicted to form the lightest hybrid supermultiplet consisting of a S-wave color-octet
$q\bar q$ pair coupled to a $J_g^{P_gC_g}=1^{+-}$ gluonic excitation. This supermultiplet was confirmed in the heavy quark sector
using lattice QCD~\cite{2012-Liu-p126-126} and the P-wave quasigluon approach~\cite{2008-Guo-p56003-56003}.
Our result supports this prediction although the extracted masses are $0.4\sim0.7$ GeV lower. A heavier hybrid supermultiplet in
Refs.~\cite{2012-Liu-p126-126, 2008-Guo-p56003-56003} contains states with $J^{PC}=0^{+-}, (1^{+-})^3, (2^{+-})^2, 3^{+-}, (0, 1, 2)^{++}$.
It is composed of a P-wave color-octet $q\bar q$ pair coupled to a gluonic field with $J_g^{P_gC_g}=1^{+-}$. In Table~\ref{table1},
we list our mass predictions for the $J^{PC}=(0, 1)^{+-}, (0, 1, 2)^{++}$ members of this excited hybrid supermultiplet.

We obtain three charmonium hybrid states with exotic quantum numbers $J^{PC}=1^{-+}, 0^{+-}, 0^{--}$ in Table~\ref{table1}.
The experimental identification of these states is considered to be a smoking gun for the existence of the gluonic degree of freedom in
QCD because they cannot mix with the conventional charmonium states. The lightest exotic charmonium hybrid state
with $J^{PC}=1^{-+}$ in the first hybrid supermultiplet is particularly important. The mass of the heaviest hybrid charmonium
with $J^{PC}=0^{--}$ is about $5.5$ GeV, which is consistent with the lattice QCD prediction~\cite{2006-Liu-p54510-54510}
within the errors. This large mass may suggest that it has highly excited gluonic structures.

In principle, conventional charmonium $c\bar c$ states can also couple to the charmonium hybrid currents with the same quantum numbers. Although a full analysis of the  mixed hybrid/charmonium scenario is beyond the scope of this work (see {\it e.g.,} Ref.~\cite{2008-Kisslinger-p982341-982341}) we can explore the qualitative effect of mixing on our mass predictions by modifying the right-hand side of \eqref{sumrule} to include a conventional charmonium ground state $m_{c\bar c}$
\begin{eqnarray}
\mathcal{L}_{k}\left(s_0, M_B^2\right)=f_X^2 m_X^{8+2k}e^{-m_X^2/M_B^2}
+f^2_{c} m_{c\bar c}^{8+2k}e^{-m_{c\bar c}^2/M_B^2}
=\int_{4m^2}^{s_0}ds\,s^k\,\rho(s)\,e^{-s/M_B^2}\,,
\label{sumrulemixing}
\end{eqnarray}
so that \eqref{mass} becomes
\begin{eqnarray}
m_X^2=\frac{\mathcal{L}_{1}\left(s_0\,, M_B^2\right)-f_c^2m_{c\bar c}^{10}e^{-m_{c\bar c}^2/M_B^2}}{\mathcal{L}_{0}\left(s_0\,, M_B^2\right) -f_c^2m_{c\bar c}^{8}e^{-m_{c\bar c}^2/M_B^2}}.
\label{massmixing}
\end{eqnarray}
By inputting the known charmonium mass $m_{c\bar c}$ and allowing $f_c$ to increase from zero (\textit{i.e.} the pure hybrid case) we can determine how the hybrid mass is influenced by mixing.  Fig.~\ref{mixfig} shows that the $1^{--}$ hybrid state mass increases as  mixing with conventional charmonium increases and similar behaviour is found for all the other non-exotic cases.  Mixing effects will thus tend to raise the mass predictions of Table~\ref{table1} implying that our results are a lower bound on the mixed hybrid mass.  A conservative estimate would be a mixed mass in the upper range of  the uncertainties given in Table~\ref{table1}.

\begin{center}
\begin{tabular}{lr}
\scalebox{0.8}{\includegraphics{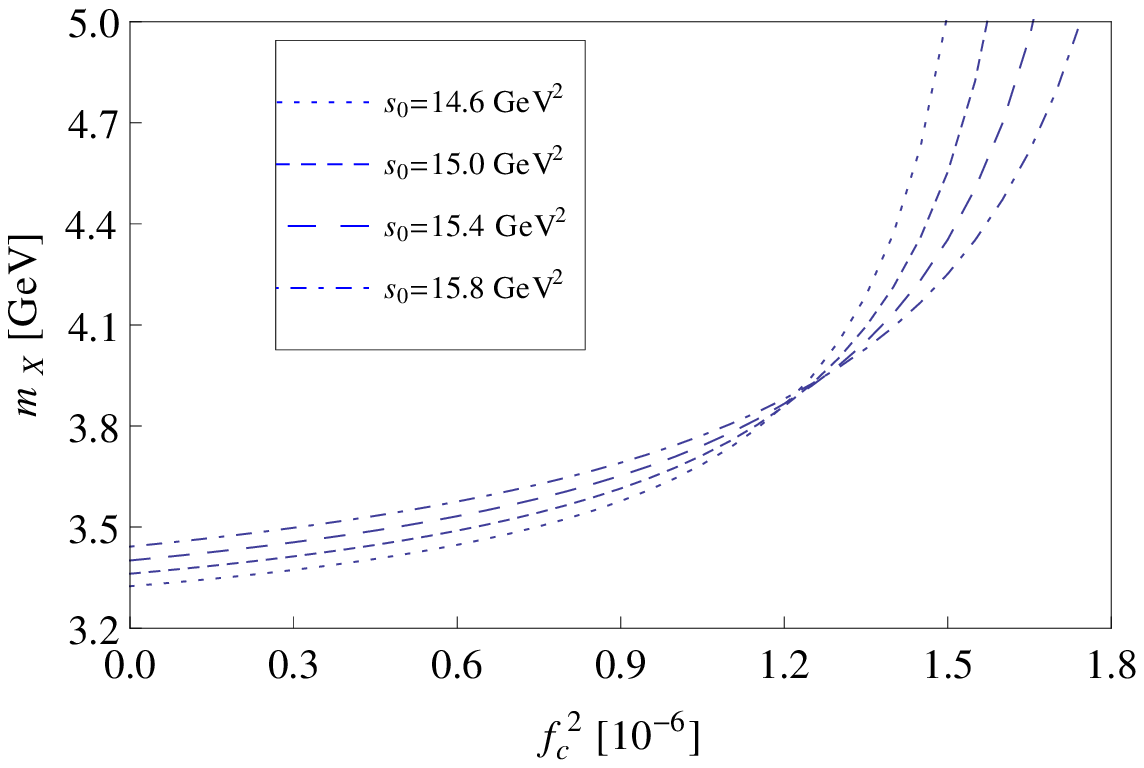}}&
\end{tabular}
\figcaption{The variations of $m_X$ with $f_c$    for $J^{PC}=1^{--}$ 
hybrid charmonium system.  The Borel scale is the central value from Table~\ref{table1} and we have chosen several values of   $s_0$  near the optimum value of Table~\ref{table1}.  }\label{mixfig}
\end{center}

The quantum numbers of $X(3872)$ are given as $J^{PC}=1^{++}$ or $2^{-+}$ in the PDG ~\cite{2012-Beringer-p10001-10001}.
Although the analysis of angular distributions favors the assignment $J^{PC}=1^{++}$~\cite{2005-Abe-p-, 2007-Abulencia-p132002-132002}, 
the $2^{-+}$ assignment is also possible~\cite{2010-AmoSanchez-p11101-11101}. Recently, the LHCb collaboration has provided strong
evidence for $J^{PC}=1^{++}$ \cite{2013-Aaij-p222001-222001}. As previously found in Ref.~\cite{2012-Harnett-p125003-125003}, the mass of the $1^{++}$ hybrid charmonium in Table~\ref{table1}
is around $5.06$ GeV, which is much higher than the mass of $X(3872)$, which seems to preclude a pure hybrid interpretation as pointed out in Ref.~\cite{2012-Harnett-p125003-125003}. The mass of the $2^{-+}$ hybrid charmonium is about $4.45$ GeV, which also precludes the $2^{-+}$ hybrid charmonium assignment for the $X(3872)$, in agreement with the LHCb result~\cite{2013-Aaij-p222001-222001}.

\begin{center}
\begin{tabular}{lr}
\scalebox{0.6}{\includegraphics{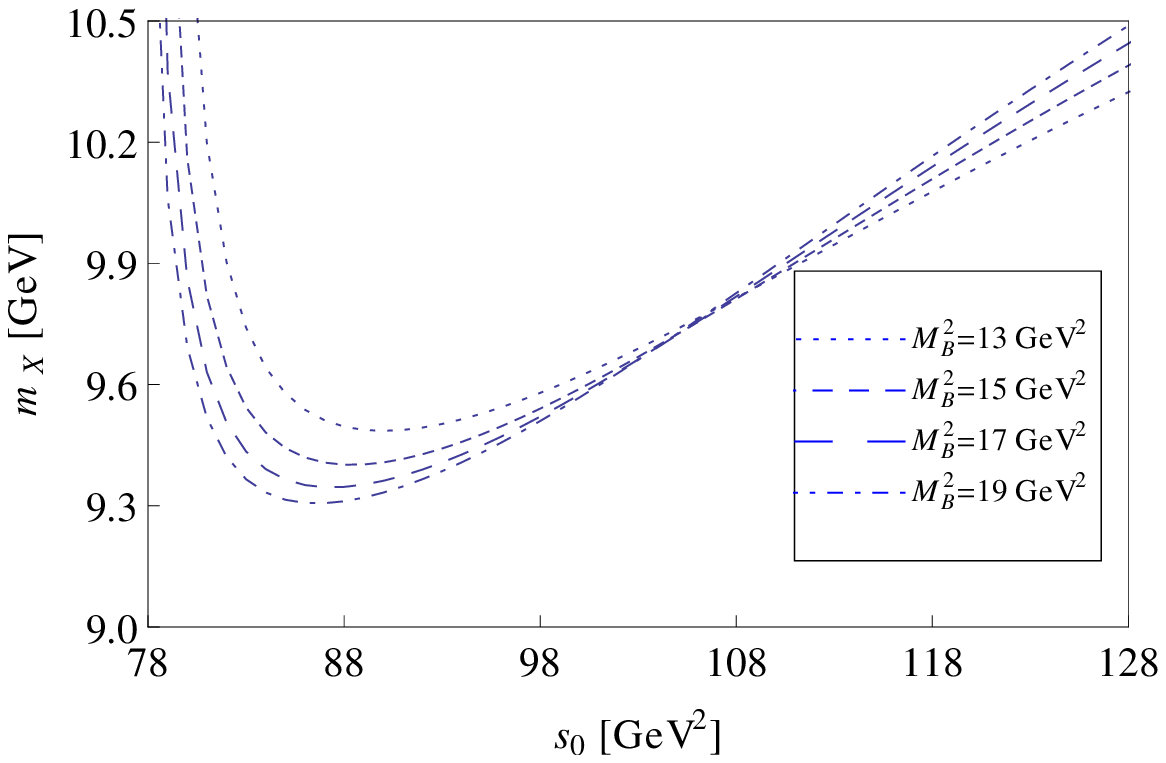}}&
\scalebox{0.6}{\includegraphics{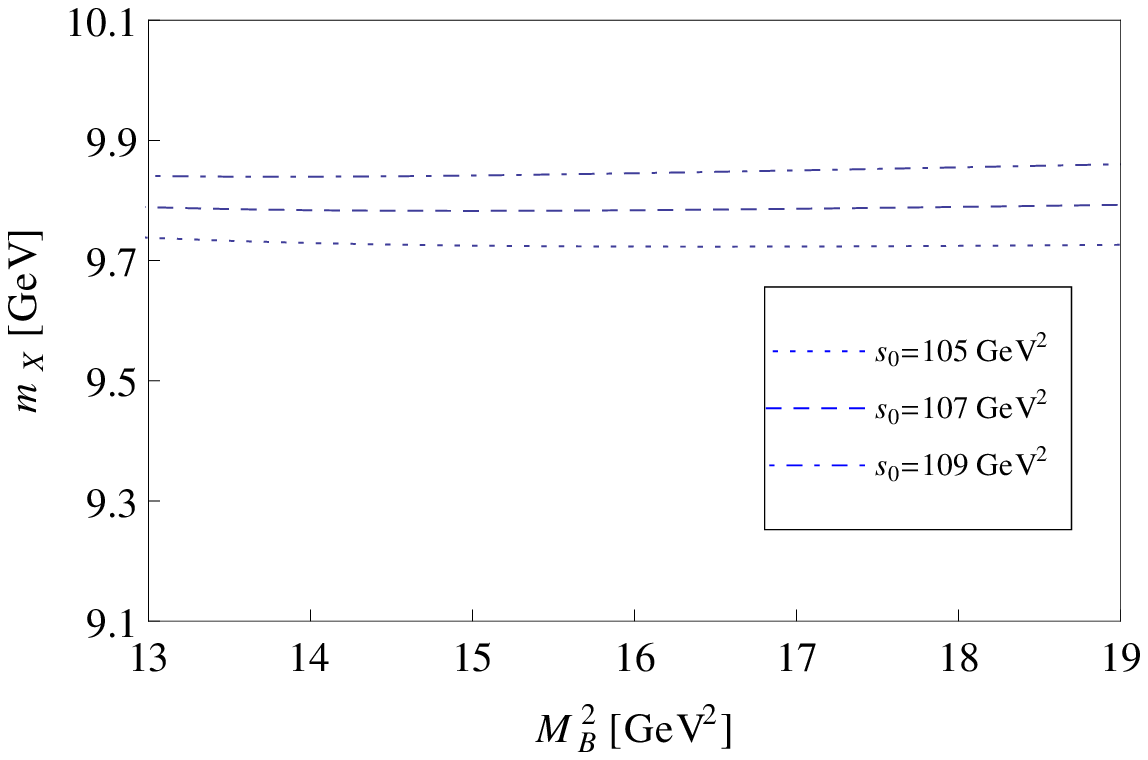}}
\end{tabular}
\figcaption{The variations of $m_X$ with $s_0$ and $M_B^2$ for $J^{PC}=1^{-+}$ hybrid bottomonium system.} \label{fig1-+bb}
\end{center}

By replacing $m_c$ with $m_b$ in the spectral densities and using the strong coupling in Eq.~(\ref{alpha_bb}), we perform the
same analysis as described above and collect the numerical results for bottomonium hybrid states in Table~\ref{table2}. Obviously the
Borel windows are enlarged compared to the hybrid charmonium systems, which means that the stabilities are better for the bottomonium
hybrid states. We show the Borel curves of the bottomonium hybrid state with $J^{PC}=1^{-+}$ in Fig.~\ref{fig1-+bb}. In Table~\ref{table2},
the masses of the four bottomonium hybrid states with $J^{PC}=(0, 1, 2)^{-+}, 1^{--}$ are about $9.7\sim 9.9$ GeV. The mass variations of 
these states are much smaller than those in the charmonium system. They form the lightest bottomonium 
supermultiplet, in complete analogy with the charmonium hybrid supermultiplet structure. In Table~\ref{table2}, the masses of $0^{-+}$ and $1^{++}$ bottomonium hybrids are again in agreement with Refs.~\cite{2012-Berg-p34002-34002,2012-Harnett-p125003-125003} and the $0^{--}$ state is still the heaviest one.

\section{SUMMARY}\label{sec:SUMMARY}
In this paper we have studied the mass spectrum of charmonium and bottomonium hybrids using QCD sum rules.
Using the hybrid operators in Eq.~(\ref{currents}), we calculate the correlation functions and the spectral densities,
confirming the perturbative and gluon condensate results of \cite{1985-Govaerts-p215-215, 1985-Govaerts-p575-575,
1987-Govaerts-p674-674} and extending them to include the dimension six condensates $\GGGb$ and $\JJb$. Agreement with
the $\GGGb$ results of Refs.~\cite{2012-Berg-p34002-34002,2012-Harnett-p125003-125003} provides an additional consistency
check on our analysis. Expressions for the correlation functions and spectral densities are given in Appendix~\ref{sec:rhos}.

The Feynman diagram in Fig.~(\ref{fig3}a) gives contributions to both the condensates $\GGGb$ and $\JJb$. The condensate $\JJb$ has not previously been calculated in hybrid sum rules. Given the crucial stabilizing contribution of the dimension six gluon condensate, we have also included the dimension six quark condensate in our analysis. The channels with $J^{PC}=0^{-+}, 0^{+-}, 1^{-+}, 1^{--}, 2^{-+}$ were unstable without the contributions of the dimension six condensates in the original QCD sum rule analysis~\cite{1985-Govaerts-p215-215, 1985-Govaerts-p575-575, 1987-Govaerts-p674-674}. However, these channels are stabilized by the effects of the dimension six condensates in our analysis. In our analysis all channels lead to stable mass sum rules. Due to the small values of the pole contribution, the Borel windows of the channels with $J^{PC}=0^{-+}, 0^{+-}, 1^{-+}, 1^{--}, 2^{-+}$ are much narrower than those of $J^{PC}=0^{--}, 0^{++}, 1^{++}, 1^{+-}, 2^{++}$.
We speculate that this explains why these channels were unstable in Refs.~\cite{1985-Govaerts-p215-215,1985-Govaerts-p575-575,
1987-Govaerts-p674-674}; the pole contributions were so small that no effective working regions for Borel
parameter existed. Although the $\JJb$ effects are very small compared to $\GGGb$, we still include its contribution to the
correlation function in this work because of the important stabilizing role of the dimension six condensates.

The sum-rule mass predictions show that for the charmonium hybrids, there are four negative-parity states with $J^{PC}=(0, 1, 2)^{-+}, 1^{--}$ which lie below 4 GeV. Their masses are about $3.4\sim 3.9$ GeV and are much lower than the other states, forming the lightest charmonium hybrid supermultiplet. Our mass prediction for the $1^{--}$ state is comparable to the result found in Ref.~\cite{2008-Kisslinger-p982341-982341}, which provides indirect support for our mass predictions for the states in the lightest supermultiplet. 
These states were predicted to form the lightest hybrid supermultiplet in lattice QCD~\cite{2012-Liu-p126-126} and the P-wave quasigluon approach~\cite{2008-Guo-p56003-56003},
consisting of a S-wave color-octet $q\bar q$ pair and a $1^{+-}$ excited gluonic field. A heavier hybrid supermultiplet is
composed of a P-wave color-octet $q\bar q$ pair coupled to a $1^{+-}$ excited gluonic field~\cite{2012-Liu-p126-126, 2008-Guo-p56003-56003}.
In our spectrum, we predict masses of five positive-parity members of this excited hybrid supermultiplet with $J^{PC}=(0, 1)^{+-}, (0, 1, 2)^{++}$.
We speculate that the supermultiplets are occupied by states of the same parity because of the behaviour of the perturbative, gluon condensate 
and dimension six condensate contributions under parity reversal. In particular, Appendix~\ref{sec:rhos} shows that the perturbative contributions are invariant and the gluon condensate and tri-gluon condensate contributions change sign under parity reversal, implying a consistent pattern in the mass hierarchy of the positive- and negative-parity states. We note that the $J^{PC}=0^{--}$ hybrid has a much higher mass in our spectrum which may suggest that it involves a different excitation of the gluonic field compare to other channels. We list the masses of the charmonium and bottomonium hybrid states in Table~\ref{table1} and Table~\ref{table2}, respectively. Finally, we have investigated the effects of mixing between non-exotic hybrids and conventional quarkonia. We find that mixing increases the mass predictions for the non-exotic hybrids given in Tables~\ref{table1} and \ref{table2}, and we conservatively estimate that the mixed states masses are in the upper range of the uncertainties given in Tables~\ref{table1} and \ref{table2}. A more detailed study of mixing between non-exotic hybrids and conventional quarkonia is left for future work. 

To date, there are many interpretations of the $X(3872)$ meson, such as a molecular state, a tetraquark state, and a charmonium hybrid.
LHCb has provided strong evidence for the $J^{PC}=1^{++}$ quantum numbers, seemingly ruling out the $2^{-+}$ assignment \cite{2013-Aaij-p222001-222001}. In our spectrum, the masses of the $1^{++}$ and $2^{-+}$ charmonium hybrids are about $5.06$ GeV and $4.45$ GeV, respectively. The $1^{++}$ (also see Ref.~\cite{2012-Harnett-p125003-125003}) and $2^{-+}$ mass predictions seem to preclude a pure charmonium hybrid interpretation of the $X(3872)$ meson.

Observation of the overpopulation of the states with $q\bar q$ quantum numbers in the charmonium and bottomonium regions, such as
the $X, Y, Z$ states, is an important signal for the existence of the heavy quark hybrids. As an experimental signal, this overpopulation complements the identification of states with exotic quantum numbers. Experiments such as BESIII, PANDA and LHCb will collect more definitive data on the
charmonium spectrum, including hybrid mesons. Our results for the mass spectra of charmonium and bottomonium hybrids in Table~\ref{table1} and Table~\ref{table2} may be useful for the future search for these fascinating exotic and non-exotic
hybrid states at these experimental facilities.

\section*{Acknowledgments}

This project was supported by the Natural Sciences and Engineering
Research Council of Canada (NSERC). S.L.Z. was supported by the
National Natural Science Foundation of China under Grants
11075004, 11021092, 11261130311 and Ministry of Science and
Technology of China (2009CB825200).




\appendix

\section{Correlation Functions and Spectral Densities}\label{sec:rhos}

In this appendix, we list the correlation functions and spectral densities of the hybrid interpolating currents in Eq.~(\ref{currents}). First, we tabulate the correlation functions obtained by the first approach mentioned in Sec~\ref{sec:QSR}. For all of the various $J^{PC}$ quantum numbers studied in this paper, the perturbative contributions can be expressed as
\begin{gather}
\Pi_{\rm pert}\left(z\right) = \frac{m^6\alpha_s}{1620\pi^3} \left(f_1\left(z\right)\phantom{}_{3}F_2\left[1\,,1\,,1\,;\frac{3}{2}\,,3\,;z\right] +f_2\left(z\right)\phantom{}_{3}F_2\left[1\,,1\,,2\,;\frac{5}{2}\,,4\,;z\right] +\frac{f_3\left(z\right)}{z} \right) \,, \quad z=\frac{q^2}{4m^2} \,,
\label{pert_contribution_to_correlation_fcn}
\end{gather}
where we have omitted terms corresponding to dispersion relation subtraction constants, $\phantom{}_{3}F_2$ denotes the generalized hypergeometric function~\cite{Bateman:1953} and the functions $f_1\left(z\right)$, $f_2\left(z\right)$ and $f_3\left(z\right)$ are polynomials in $z$ that are tabulated in Table~\ref{pert_and_gluon_cond_correlation_fcn_polynomials} for each heavy hybrid channel studied in this paper. For the dimension four gluon condensate $\langle \alpha_s GG\rangle$, all contributions can be expressed as
\begin{gather}
\Pi_{\rm GG}\left(z\right) = \frac{m^2\GGb}{54\pi} f_4\left(z\right)\phantom{}_{2}F_1\left[1\,,1\,;\frac{3}{2}\,;z\right] \,,
\label{gluon_cond_contribution_to_correlation_fcn}
\end{gather}
where we have again omitted terms corresponding to dispersion relation subtraction constants, $\phantom{}_{2}F_1$ denotes the Gauss hypergeometric function~\cite{Bateman:1953}, $z$ is defined as in~\eqref{pert_contribution_to_correlation_fcn} and the function $f_4\left(z\right)$ are polynomials in $z$ that are tabulated in Table~\ref{pert_and_gluon_cond_correlation_fcn_polynomials} for each heavy hybrid channel studied in this paper.

\begin{center}
\begin{tabular}{ccccc}
\hlinewd{.8pt}
$J^{PC}$ & $f_1\left(z\right)$ & $f_2\left(z\right)$ & $f_3\left(z\right)$ & $f_4\left(z\right)$ \\
\hline
$1^{++}$ & $36 \left(-5+8 z-15 z^2+12 z^3\right)$ & $ 4 z \left(-5+7 z-96 z^2+24 z^3\right) $ & $0$ &  $ -2 z (1+2 z) $ \\

$0^{-+}$ & $ 54 (-10 + 31 z - 25 z^2 + 4 z^3)$ & $ 6 z \left(-10+29 z+8 z^2+8 z^3\right) $ & $0$ &  $ 3 z (1+2 z) $ \\

$1^{--}$ & $ 9 (5 - 53 z + 48 z^3) $ & $ z (5 - 52 z - 504 z^2 + 96 z^3) $ & $0$ &  $ -z (-7+4 z) $ \\

$0^{+-}$ & $ 27 (5 - 23 z + 10 z^2 + 8 z^3) $ & $ 3 z (5 - 22 z - 104 z^2 + 16 z^3) $ & $0$ &  $ 6 (-1+z) z $ \\





$2^{-+}$ & $ \frac{3}{16 z} \left(-35-49 z-60 z^2-496 z^3+640 z^4\right) $ & $ \frac{1}{48} \left(-35-56 z-72 z^2-5632 z^3+1280 z^4\right)$  &  $\frac{105}{16} $ & $ -\frac{3}{2}z$\\
\hline
\end{tabular}
\tabcaption{Polynomials $f_1\left(z\right)$, $f_2\left(z\right)$, $f_3\left(z\right)$  for the perturbative contribution~\eqref{pert_contribution_to_correlation_fcn} and $f_4\left(z\right)$ for the gluon condensate contribution~\eqref{gluon_cond_contribution_to_correlation_fcn}. Note that for $J^{PC}=1^{-+}\,,0^{++}\,,1^{+-}\,,0^{--}\,,2^{++}$ omitted above, the polynomials $f_1\left(z\right)$, $f_2\left(z\right)$, $f_3\left(z\right)$ are identical to those for $J^{PC}=1^{++}\,,0^{-+}\,,1^{--}\,,0^{+-}\,,2^{-+}$, whereas the polynomials $f_4\left(z\right)$ differ by an overall minus sign.} \label{pert_and_gluon_cond_correlation_fcn_polynomials}
\end{center}

Now we consider contributions from the dimension six gluon condensate, $\langle g^3 GGG\rangle$. For convenience, we tabulate contributions from the left diagram and right diagram in~Fig.~\ref{fig3} seperately. The left diagram, which arises from the vacuum expectation value $\langle DDG \rangle$ is denoted as $\Pi^a_{\rm GGG}$, while the right diagram is denoted as $\Pi^b_{\rm GGG}$. All contributions to $\Pi^a_{\rm GGG}$ can be expressed as

\begin{gather}
\Pi^a_{\rm GGG}\left(z\right) = \frac{\GGGb}{1152\pi^2}\frac{1}{(z-1)^2} \left(f_5\left(z\right) +f_6\left(z\right)\phantom{}_{2}F_1\left[1\,,1\,;\frac{3}{2}\,;z\right] \right) \,.
\label{g3a_cond_contribution_to_correlation_fcn}
\end{gather}
Similarly, all contributions to $\Pi^b_{\rm GGG}$ can be expressed as
\begin{gather}
\Pi^b_{\rm GGG}\left(z\right) = \frac{\GGGb}{1152\pi^2}\frac{1}{z-1} \left(f_7\left(z\right) +f_8\left(z\right)\phantom{}_{2}F_1\left[1\,,1\,;\frac{3}{2}\,;z\right] \right) \,, 
\label{g3b_cond_contribution_to_correlation_fcn}
\end{gather}
where in Eqns.~\eqref{g3a_cond_contribution_to_correlation_fcn} and~\eqref{g3b_cond_contribution_to_correlation_fcn} we have omitted terms corresponding to dispersion relation subtraction constants, all notations are identical to those in~\eqref{gluon_cond_contribution_to_correlation_fcn} and the functions $f_5\left(z\right)$, $f_6\left(z\right)$, $f_7\left(z\right)$ and $f_8\left(z\right)$ are polynomials in $z$ that are tabulated in Table~\ref{g3ab_cond_correlation_fcn_polynomials} for each heavy hybrid channel studied in this paper.

\begin{center}
\begin{tabular}{ccccc}
\hlinewd{.8pt}
$J^{PC}$ & $f_5\left(z\right)$ & $f_6\left(z\right)$ & $f_7\left(z\right)$ & $f_8\left(z\right)$  \\
\hline
$1^{++}$ & $ 51-138 z+81 z^2$ & $ -2 z \left(2-9 z+6 z^2\right) $ & $ 27-51 z $  & $ 4 z (-1+3 z)$  \\

$0^{-+}$ & $ -48+150 z-93 z^2 $ & $ -3+6 z-18 z^2+12 z^3 $ & $ -27+63 z $  &  $ -12 z^2 $ \\

$1^{--}$ & $ 78-156 z+81 z^2$ & $ -3-4 z+18 z^2-12 z^3 $ & $63-51 z $  &  $4 z (-4+3 z) $ \\

$0^{+-}$ & $ -3 (-1+z) (-25+31 z) $ & $ 6 (-1+z) z (-1+2 z) $ & $ 63 (-1+z)$  &  $-12 (-1 + z) z $ \\

$2^{-+}$ & $\frac{1}{6} \left(5-46 z+32 z^2\right) $ & $ \frac{1}{2} $ &  $0$ & $0$  \\


$1^{-+}$ & $ -3 \left(-3-6 z+7 z^2\right)$ & $ $ & $ 3 (3+5 z)$  \\


$0^{++}$ & $ -12-30 z+33 z^2$ & $ $ & $ -9-27 z $ \\

$1^{+-}$ & $ -3 \left(6-12 z+7 z^2\right)$ & $ $ & $ -3 (9-5 z)$  \\


$0^{--}$ & $ -3 (5-11 z) (-1+z)$ & $  $ & $ -27 (-1+z)$ & $  $ \\

$2^{++}$ & $\frac{1}{6} \left(-13+62 z-40 z^2\right) $ & $ $ &  $0$ & $ $  \\
\hline
\hlinewd{.8pt}
\end{tabular}
\tabcaption{Polynomials $f_5\left(z\right)$, $f_5\left(z\right)$ for the contribution~\eqref{g3a_cond_contribution_to_correlation_fcn} and  $f_7\left(z\right)$, $f_8\left(z\right)$ for the contribution~\eqref{g3b_cond_contribution_to_correlation_fcn} to the dimension six gluon condensate. Note that the polynomials $f_6\left(z\right)$ and $f_8\left(z\right)$ for $J^{PC}=1^{-+}\,,0^{++}\,,1^{+-}\,,0^{--}\,,2^{++}$ only differ from those for $J^{PC}=1^{++}\,,0^{-+}\,,1^{--}\,,0^{+-}\,,2^{-+}$ by an overall minus sign.} \label{g3ab_cond_correlation_fcn_polynomials}
\end{center}
We do not include explicit expressions for the dimension six quark condensate correlation function, although integral representations for these contributions are included in the following.

Note also that the imaginary parts corresponding to the perturbative~\eqref{pert_contribution_to_correlation_fcn}, dimension four gluon condensate~\eqref{gluon_cond_contribution_to_correlation_fcn} and dimension six gluon condensate~\eqref{g3a_cond_contribution_to_correlation_fcn},~\eqref{g3b_cond_contribution_to_correlation_fcn} correlation functions can be extracted through analytic continuations of the hypergeometric functions in these expressions. This is precisely what was done in Refs~\cite{2012-Berg-p34002-34002,2012-Harnett-p125003-125003}. Given the universal forms of these hypergeometric functions for all $J^{PC}$ channels, we could construct tables similar to those above for each imaginary part. However, we have not chosen to do so. Instead, we have listed explicit expressions for these imaginary parts in the form of spectral densities in order to emphasize the equivalence of the two approaches that we have taken to calculate the Wilson coefficients. In what follows, all spectral densities will be 
given in terms of a closed form expression and an integral representation, and we have verified that these are numerically identical.

Using the second approach mentioned in Sec.~\ref{sec:QSR}, we calculate the spectral densities up to dimension six:
\begin{eqnarray}
\rho(s)=\rho^{pert}(s)+\rho^{\GGa}(s)+\rho^{\GGGa}(s)+\rho^{\JJa}(s).
\end{eqnarray}
The integration limits in the following expressions are:
\begin{eqnarray}
\nonumber
\alpha_{max}&=&\frac{1+\sqrt{1-4m^2/s}}{2},\hspace{1cm}
\alpha_{min}=\frac{1-\sqrt{1-4m^2/s}}{2}\\
\beta_{max}&=&1-\alpha,\hspace{2.8cm} \beta_{min}=\frac{\alpha
m^2}{\alpha s-m^2}.
\end{eqnarray}

\begin{itemize}
\item For $J^{PC}=1^{-+}$:
{\allowdisplaybreaks
\begin{eqnarray}
\nonumber \rho^{pert}(s)&=&-\dab\frac{g^2(1-\alpha-\beta)}{96\pi^4}\Big\{\frac{(\alpha+\beta)^3(9\alpha^2+12\alpha\beta-9\alpha+3\beta^2-7\beta)m^6}
{\alpha^2\beta^3}
\non &&-\frac{3(\alpha+\beta)^2(20\alpha^2+24\alpha\beta-25\alpha+4\beta^2-13\beta+5)m^4s}
{\alpha\beta^2}
\non &&+\frac{3(\alpha+\beta)(35\alpha^2+40\alpha\beta-49\alpha+5\beta^2-23\beta+14)m^2s^2}
{\beta}
\non&&-\alpha(54\alpha^2+60\alpha\beta-81\alpha+6\beta^2-37\beta+27)s^3\Big\},
\non
&=&\frac{m^6 \alpha_s}{180 \pi ^3  z^2} \Biggl(15 \log{\left[\sqrt{z}+\sqrt{z-1}\right]}  \left(1-3 z+16 z^3\right) \Biggr.
\non
&\Biggl.&\qquad\qquad +\sqrt{z}\sqrt{z-1}\left(15-35 z-22 z^2-216 z^3+48 z^4\right)\Biggr)\,, \quad z=\frac{s}{4m^2}\,,
\non
\rho^{\GGa}(s)&=&\frac{\GGb(2m^2+s)}{36\pi}\sqrt{1-4m^2/s},
\non
\rho^{\GGGa}(s)&=&\frac{\GGGb}{192\pi^2}\int^1_0dx\Big\{\frac{(1+4x-4x^2)m^4}{x(1-x)^3}\delta'(s-\tilde{m}^2)
-\frac{(1+4x-8x^2+2x^3)m^2}{x(1-x)^2}\delta(s-\tilde{m}^2)
\non&&-(1+2x)\theta(s-\tilde{m}^2)\Big\}
\non
&=&-\frac{\GGGb m^2s}{96\pi^2(s-4m^2)^2}\sqrt{1-4m^2/s},
\\
\rho^{\JJa}(s)&=&\frac{\JJb}{144\pi^2}\int^1_0dx\Big\{\frac{4m^4}{(1-x)^2}\delta'(s-\tilde{m}^2)
+\frac{(1+2x)m^2}{1-x}\delta(s-\tilde{m}^2)+3x\theta(s-\tilde{m}^2)\Big\}
\non&=&
\frac{\JJb (3s^2-18m^2s+16m^4)}{288\pi^2(s-4m^2)^2}\sqrt{1-4m^2/s}, \label{SP1-+}
\non
\mathcal{L}_0(M_B^2,s_0)&=&\int_{4m^2}^{s_0} \left[\rho^{pert}(s)+\rho^{\GGa}(s)\right]e^{-s/M_B^2}ds+\lim_{\eta\to 0^+}\Bigg\{
\non
&&\int_{4m^2(1+\eta)}^{s_0}\left[\rho^{\GGGa}(s)+\rho^{\JJa}(s)\right]e^{-\frac{s}{M_B^2}}ds
+\frac{4m^2}{\sqrt{\eta}}\left[\frac{\GGGb}{192\pi^2}+\frac{\JJb}{288\pi^2}\right]e^{-\frac{4m^2}{M_B^2}}
\Bigg\},
\end{eqnarray}
}
\item For $J^{PC}=0^{++}$:
{\allowdisplaybreaks
\begin{eqnarray}
\nonumber \rho^{pert}(s)&=&\dab\frac{g^2(1-\alpha-\beta)}{96\pi^4}\Big\{\frac{(\alpha+\beta)^3(9\alpha^2+12\alpha\beta-9\alpha+3\beta^2-7\beta)m^6}
{\alpha^2\beta^3}
\non &&-\frac{3(\alpha+\beta)(12\alpha^3+36\alpha^2\beta-15\alpha^2+3\alpha-36\alpha\beta+36\alpha\beta^2+12\beta^3-21\beta^2+7\beta)m^4s}
{\alpha\beta^2}
\non &&+\frac{3(1-\alpha-\beta)(2\alpha+5\beta)(2-5\alpha-5\beta)m^2s^2}
{\beta}
\non&&-\alpha(18\alpha^2+60\alpha\beta-27\alpha+42\beta^2-49\beta+9)s^3\Big\},
\non
&=&\frac{m^6 \alpha_s}{120 \pi ^3  z^2} \Biggl(-15 \log{\left[\sqrt{z}+\sqrt{z-1}\right]}  \left(-2+9 z-16 z^2+16 z^3\right) \Biggr.
\non
&\Biggl.&\qquad\qquad +\sqrt{z}\sqrt{z-1}\left(30-115 z+166 z^2+8 z^3+16 z^4\right)\Biggr)\,, \quad z=\frac{s}{4m^2}\,,
\non
\rho^{\GGa}(s)&=&-\frac{\GGb(2m^2+s)}{24\pi}\sqrt{1-4m^2/s},
\non
\rho^{\GGGa}(s)&=&-\frac{\GGGb}{192\pi^2}\int^1_0dx\Big\{\frac{2(1+2x-2x^2)m^4}{x(1-x)^3}\delta'(s-\tilde{m}^2)
-\frac{2(1+x+2x^2)m^2}{x(1-x)}\delta(s-\tilde{m}^2)
\non&&-(1+2x)\theta(s-\tilde{m}^2)\Big\}
\non
&=&\frac{\GGGb m^2(s^2-4m^2s+8m^4)}{32\pi^2(s-4m^2)^2s}\sqrt{1-4m^2/s},
\non
\rho^{\JJa}(s)&=&-\frac{\JJb}{288\pi^2}\int^1_0dx\Big\{\frac{(3+6x-16x^2+8x^3)m^4}{x(1-x)^4}\delta'(s-\tilde{m}^2)
\\&&-\frac{2(4x^3-10x^2+9x-2)m^2}{(1-x)^3}\delta(s-\tilde{m}^2)
-4x\theta(s-\tilde{m}^2)\Big\}
\non&=&
-\frac{\JJb (s^3-6m^2s^2+8m^4s-32m^6)}{96\pi^2(s-4m^2)^2s}\sqrt{1-4m^2/s},
\non
\mathcal{L}_0(M_B^2,s_0)&=&\int_{4m^2}^{s_0} \left[\rho^{pert}(s)+\rho^{\GGa}(s)\right]e^{-s/M_B^2}ds+\lim_{\eta\to 0^+}\Bigg\{
\non
&&\int_{4m^2(1+\eta)}^{s_0}\left[\rho^{\GGGa}(s)+\rho^{\JJa}(s)\right]e^{-\frac{s}{M_B^2}}ds
-\frac{4m^2}{\sqrt{\eta}}\left[\frac{\GGGb}{128\pi^2}+\frac{\JJb}{96\pi^2}\right]e^{-\frac{4m^2}{M_B^2}}
\Bigg\},
\end{eqnarray}
}

\item For $J^{PC}=1^{++}$:
{\allowdisplaybreaks
\begin{eqnarray}
\nonumber \rho^{pert}(s)&=&\rho^{pert}_{1^{-+}}(s),
\non
\rho^{\GGa}(s)&=&-\rho^{\GGa}_{1^{-+}}(s),
\non
\rho^{\GGGa}(s)&=&-\rho^{\GGGa}_{1^{-+}}(s),
\\
\rho^{\JJa}(s)&=&-2\rho^{\JJa}_{1^{-+}}(s),
\non
\mathcal{L}_0(M_B^2,s_0)&=&\int_{4m^2}^{s_0} \left[\rho^{pert}(s)+\rho^{\GGa}(s)\right]e^{-s/M_B^2}ds+\lim_{\eta\to 0^+}\Bigg\{
\non
&&\int_{4m^2(1+\eta)}^{s_0}\left[\rho^{\GGGa}(s)+\rho^{\JJa}(s)\right]e^{-\frac{s}{M_B^2}}ds
-\frac{4m^2}{\sqrt{\eta}}\left[\frac{\GGGb}{192\pi^2}+\frac{\JJb}{144\pi^2}\right]e^{-\frac{4m^2}{M_B^2}}
\Bigg\},
\end{eqnarray}
}
\item For $J^{PC}=0^{-+}$:
{\allowdisplaybreaks
\begin{eqnarray}
\nonumber \rho^{pert}(s)&=&\rho^{pert}_{0^{++}}(s),
\non
\rho^{\GGa}(s)&=&-\rho^{\GGa}_{0^{++}}(s),
\non
\rho^{\GGGa}(s)&=&-\rho^{\GGGa}_{0^{++}}(s),
\non
\rho^{\JJa}(s)&=&\frac{\JJb}{288\pi^2}\int^1_0dx\Big\{\frac{(3-8x^2+4x^3)m^4}{x(1-x)^4}\delta'(s-\tilde{m}^2)
-\frac{2(2x^3-5x^2+6x-4)m^2}{(1-x)^3}\delta(s-\tilde{m}^2)
\\ &&-2x\theta(s-\tilde{m}^2)\Big\}
\non
&=&\frac{\JJb (s^3-6m^2s^2+8m^4s-8m^6)}{48\pi^2(s-4m^2)^2s}\sqrt{1-4m^2/s},
\non
\mathcal{L}_0(M_B^2,s_0)&=&\int_{4m^2}^{s_0} \left[\rho^{pert}(s)+\rho^{\GGa}(s)\right]e^{-s/M_B^2}ds+\lim_{\eta\to 0^+}\Bigg\{
\non
&&\int_{4m^2(1+\eta)}^{s_0}\left[\rho^{\GGGa}(s)+\rho^{\JJa}(s)\right]e^{-\frac{s}{M_B^2}}ds
+\frac{4m^2}{\sqrt{\eta}}\left[\frac{\GGGb}{128\pi^2}+\frac{\JJb}{192\pi^2}\right]e^{-\frac{4m^2}{M_B^2}}
\Bigg\},
\end{eqnarray}
}

\item For $J^{PC}=1^{+-}$:
{\allowdisplaybreaks
\begin{eqnarray}
\nonumber \rho^{pert}(s)&=&-\dab\frac{g^2(1-\alpha-\beta)}{96\pi^4}\Big\{\frac{(\alpha+\beta)^3(9\alpha^2+12\alpha\beta-9\alpha+3\beta^2-\beta)m^6}
{\alpha^2\beta^3}
\non &&-\frac{3(\alpha+\beta)^2(20\alpha^2+24\alpha\beta-25\alpha+4\beta^2-9\beta+5)m^4s}
{\alpha\beta^2}
\non &&+\frac{3(\alpha+\beta)(35\alpha^2+40\alpha\beta-49\alpha+5\beta^2-21\beta+14)m^2s^2}
{\beta}
\non&&-\alpha(54\alpha^2+60\alpha\beta-81\alpha+6\beta^2-37\beta+27)s^3
\Big\},
\non
&=&\frac{m^6 \alpha_s}{720 \pi ^3  z^2} \Biggl(15 \log{\left[\sqrt{z}+\sqrt{z-1}\right]}  \left(-1+12 z-48 z^2+128 z^3\right) \Biggr.
\non
&\Biggl.&\qquad\qquad +\sqrt{z}\sqrt{z-1}\left(-15+170 z-608 z^2-1104 z^3+192 z^4\right)\Biggr)\,, \quad z=\frac{s}{4m^2}\,,
\non
\rho^{\GGa}(s)&=&\frac{\GGb(s-7m^2)}{36\pi}\sqrt{1-4m^2/s},
\non
\rho^{\GGGa}(s)&=&\frac{\GGGb}{192\pi^2}\int^1_0dx\Big\{\frac{(1-8x+8x^2)m^4}{x(1-x)^3}\delta'(s-\tilde{m}^2)
-\frac{(1-8x+4x^2+2x^3)m^2}{x(1-x)^2}\delta(s-\tilde{m}^2)
\non&&-(1+2x)\theta(s-\tilde{m}^2)\Big\}
\non
&=&\frac{m^2\GGGb}{96\pi^2}\frac{5s^2-24m^2s+24m^4}{(s-4m^2)^2s}\sqrt{1-4m^2/s},
\\
\rho^{\JJa}(s)&=&-\frac{\JJb}{144\pi^2}\int^1_0dx\Big\{\frac{4m^4}{(1-x)^2}\delta'(s-\tilde{m}^2)
+\frac{(1-4x)m^2}{1-x}\delta(s-\tilde{m}^2)-3x\theta(s-\tilde{m}^2)\Big\}
\non&=&
\frac{\JJb}{288\pi^2}\frac{3s^3-18m^2s^2+8m^4s+96m^6}{(s-4m^2)^2s}\sqrt{1-4m^2/s},
\non
\mathcal{L}_0(M_B^2,s_0)&=&\int_{4m^2}^{s_0} \left[\rho^{pert}(s)+\rho^{\GGa}(s)\right]e^{-s/M_B^2}ds+\lim_{\eta\to 0^+}\Bigg\{
\non
&&\int_{4m^2(1+\eta)}^{s_0}\left[\rho^{\GGGa}(s)+\rho^{\JJa}(s)\right]e^{-\frac{s}{M_B^2}}ds
-\frac{4m^2}{\sqrt{\eta}}\left[\frac{\GGGb}{384\pi^2}+\frac{\JJb}{288\pi^2}\right]e^{-\frac{4m^2}{M_B^2}}
\Bigg\},
\end{eqnarray}
}
\item For $J^{PC}=0^{--}$:
{\allowdisplaybreaks
\begin{eqnarray}
\nonumber \rho^{pert}(s)&=&\dab\frac{g^2(1-\alpha-\beta)}{96\pi^4}\Big\{\frac{(\alpha+\beta)^3(9\alpha^2+12\alpha\beta-9\alpha+3\beta^2-\beta)m^6}
{\alpha^2\beta^3}
\non &&-\frac{3(\alpha+\beta)(12\alpha^3+36\alpha^2\beta-15\alpha^2+3\alpha-24\alpha\beta+36\alpha\beta^2+12\beta^3-9\beta^2-\beta)m^4s}
{\alpha\beta^2}
\non &&+\frac{3(15\alpha^3+55\alpha^2\beta-21\alpha^2+6\alpha-46\alpha\beta+65\alpha\beta^2+25\beta^3-25\beta^2+2\beta)m^2s^2}
{\beta}
\non&&-\alpha(18\alpha^2+60\alpha\beta-27\alpha+42\beta^2-49\beta+9)s^3\Big\},
\non
&=&\frac{m^6 \alpha_s}{240 \pi ^3  z^2} \Biggl(15 \log{\left[\sqrt{z}+\sqrt{z-1}\right]}  \left(-1+6 z-16 z^2+32 z^3\right) \Biggr.
\non
&\Biggl.&\qquad\qquad +\sqrt{z}\sqrt{z-1} \left(-15+80 z-188 z^2-224 z^3+32 z^4\right)\Biggr)\,, \quad z=\frac{s}{4m^2}\,,
\non
\rho^{\GGa}(s)&=&-\frac{\GGb(s-4m^2)}{24\pi}\sqrt{1-4m^2/s},
\non
\rho^{\GGGa}(s)&=&-\frac{\GGGb}{192\pi^2}\int^1_0dx\Big\{\frac{2(1-2x)^2m^4}{x(1-x)^3}\delta'(s-\tilde{m}^2)
-\frac{2(1-5x+2x^2)m^2}{x(1-x)}\delta(s-\tilde{m}^2)
\non&&-(1+2x)\theta(s-\tilde{m}^2)\Big\}
\non
&=&-\frac{m^2\GGGb}{32\pi^2(s-4m^2)}\sqrt{1-4m^2/s},
\non
\rho^{\JJa}(s)&=&\frac{\JJb}{288\pi^2}\int^1_0dx\Big\{\frac{(8x^3-16x^2+8x-1)m^4}{x(1-x)^4}\delta'(s-\tilde{m}^2)
\\&&+\frac{2(8x^3-22x^2+22x-7)m^2}{(1-x)^3}\delta(s-\tilde{m}^2)
+4x\theta(s-\tilde{m}^2)\Big\}
\non
&=&-\frac{\JJb}{96\pi^2}\frac{s^2-2m^2s+8m^4}{(s-4m^2)s}\sqrt{1-4m^2/s},
\non
\mathcal{L}_0(M_B^2,s_0)&=&\int_{4m^2}^{s_0} \left[\rho^{pert}(s)+\rho^{\GGa}(s)\right]e^{-\frac{s}{M_B^2}}ds+
\lim_{\eta\to 0^+}\int_{4m^2(1+\eta)}^{s_0}\left[\rho^{\GGGa}(s)+\rho^{\JJa}(s)\right]e^{-\frac{s}{M_B^2}}ds,
\end{eqnarray}
}

\item For $J^{PC}=1^{--}$:
{\allowdisplaybreaks
\begin{eqnarray}
\nonumber \rho^{pert}(s)&=&\rho^{pert}_{1^{+-}}(s),
\non
\rho^{\GGa}(s)&=&-\rho^{\GGa}_{1^{+-}}(s),
\non
\rho^{\GGGa}(s)&=&-\rho^{\GGGa}_{1^{+-}}(s),
\\
\rho^{\JJa}(s)&=&-\frac{\JJb}{288\pi^2}\int^1_0dx\Big\{\frac{(3-16x+16x^2)m^4}{x(1-x)^3}\delta'(s-\tilde{m}^2)
+\frac{(2x^2-4x+5)m^2}{(1-x)^2}\delta(s-\tilde{m}^2)\Big\}
\non
&=&-\frac{\JJb}{144\pi^2}\frac{3s^3-18m^2s^2+20m^4s+24m^6}{(s-4m^2)^2s}\sqrt{1-4m^2/s},
\non
\mathcal{L}_0(M_B^2,s_0)&=&\int_{4m^2}^{s_0} \left[\rho^{pert}(s)+\rho^{\GGa}(s)\right]e^{-s/M_B^2}ds+\lim_{\eta\to 0^+}\Bigg\{
\non
&&\int_{4m^2(1+\eta)}^{s_0}\left[\rho^{\GGGa}(s)+\rho^{\JJa}(s)\right]e^{-\frac{s}{M_B^2}}ds
+\frac{4m^2}{\sqrt{\eta}}\left[\frac{\GGGb}{384\pi^2}+\frac{\JJb}{576\pi^2}\right]e^{-\frac{4m^2}{M_B^2}}
\Bigg\},
\end{eqnarray}
}
\item For $J^{PC}=0^{+-}$:
{\allowdisplaybreaks
\begin{eqnarray}
\nonumber \rho^{pert}(s)&=&\rho^{pert}_{0^{--}}(s),
\non
\rho^{\GGa}(s)&=&-\rho^{\GGa}_{0^{--}}(s),
\non
\rho^{\GGGa}(s)&=&-\rho^{\GGGa}_{0^{--}}(s),
\non
\rho^{\JJa}(s)&=&-\frac{\JJb}{288\pi^2}\int^1_0dx\Big\{\frac{(16x^3-20x^2+4x+1)m^4}{x(1-x)^4}\delta'(s-\tilde{m}^2)
\\&&-\frac{2(2x^3-7x^2+10x-4)m^2}{(1-x)^3}\delta(s-\tilde{m}^2)
+2x\theta(s-\tilde{m}^2)\Big\}
\non
&=&\frac{\JJb}{48\pi^2}\frac{s^2-2m^2s-4m^4}{(s-4m^2)s}\sqrt{1-4m^2/s},
\non
\mathcal{L}_0(M_B^2,s_0)&=&\int_{4m^2}^{s_0} \left[\rho^{pert}(s)+\rho^{\GGa}(s)\right]e^{-\frac{s}{M_B^2}}ds+
\lim_{\eta\to 0^+}\int_{4m^2(1+\eta)}^{s_0}\left[\rho^{\GGGa}(s)+\rho^{\JJa}(s)\right]e^{-\frac{s}{M_B^2}}ds,
\end{eqnarray}
}
\item For $J^{PC}=2^{-+}$:
{\allowdisplaybreaks
\begin{eqnarray}
\nonumber \rho^{pert}(s)&=&-\dab\frac{g^2(1-\alpha-\beta)}{192\pi^4}\Big\{\frac{(\alpha+\beta)^3(6\alpha^2+7\alpha\beta-6\alpha+\beta^2-3\beta)m^6}
{\alpha^2\beta^3}
\non &&-\frac{6(\alpha+\beta)^2(8\alpha^2-10\alpha+9\alpha\beta+\beta^2-5\beta+2)m^4s}
{\alpha\beta^2}
\non &&+\frac{9(\alpha+\beta)(10\alpha^2-14\alpha+11\alpha\beta-7\beta+\beta^2+4)m^2s^2}
{\alpha\beta}
\non&&-4\alpha(12\alpha^2-18\alpha+13\alpha\beta-9\beta+\beta^2+6)s^3\Big\},
\non
&=&\frac{m^6 \alpha_s}{34560 \pi ^3  z^3} \Biggl( 15\log{\left[\sqrt{z}+\sqrt{z-1}\right]}  \left(7-256 z^3+1152 z^4\right) \Biggr.
\non &\Biggl.&\qquad\qquad +\sqrt{z}\sqrt{z-1}\left(105+70 z+56 z^2-3792 z^3-12544 z^4+2560 z^5\right)\Biggr)\,, \quad z=\frac{s}{4m^2}\,,
\non
\rho^{\GGa}(s)&=&\frac{\GGb m^2}{24\pi}\sqrt{1-4m^2/s},
\non
\rho^{\GGGa}(s)&=&\frac{\GGGb}{96\pi^2}\int^1_0dx\frac{m^4}{(1-x)^2}\delta'(s-\tilde{m}^2)
\non
&=&-\frac{\GGGb m^6}{24\pi^2(s-4m^2)^2s}\sqrt{1-4m^2/s},
\non
\rho^{\JJa}(s)&=&-\frac{\JJb}{576\pi^2}\int^1_0dx\Big\{\frac{2(4x-3)m^4}{(1-x)^3}\delta'(s-\tilde{m}^2)
+\frac{(4x^2-6x+1)m^2}{(1-x)^2}\delta(s-\tilde{m}^2)
\\&&-(4x^2-2x)\theta(s-\tilde{m}^2)\Big\}
\non
&=&-\frac{\JJb}{864\pi^2}\frac{s^3-6m^2s^2+6m^4s+32m^6}{(s-4m^2)^2s}\sqrt{1-4m^2/s}, \label{SP2-+}
\non
\mathcal{L}_0(M_B^2,s_0)&=&\int_{4m^2}^{s_0} \left[\rho^{pert}(s)+\rho^{\GGa}(s)\right]e^{-s/M_B^2}ds+
\lim_{\eta\to 0^+}\Bigg\{
\non
&&\int_{4m^2(1+\eta)}^{s_0}\left[\rho^{\GGGa}(s)+\rho^{\JJa}(s)\right]e^{-\frac{s}{M_B^2}}ds
+\frac{4m^2}{\sqrt{\eta}}\left[\frac{\GGGb}{768\pi^2}+\frac{\JJb}{1152\pi^2}\Bigg]e^{-\frac{4m^2}{M_B^2}}
\right\},
\end{eqnarray}
}

\item For $J^{PC}=2^{++}$:
{\allowdisplaybreaks
\begin{eqnarray}
\nonumber \rho^{pert}(s)&=&\rho^{pert}_{2^{-+}}(s),
\non
\rho^{\GGa}(s)&=&-\rho^{\GGa}_{2^{-+}}(s),
\non
\rho^{\GGGa}(s)&=&-\rho^{\GGGa}_{2^{-+}}(s),
\non
\rho^{\JJa}(s)&=&-\frac{\JJb}{576\pi^2}\int^1_0dx\Big\{\frac{2(3-2x)m^4}{(1-x)^3}\delta'(s-\tilde{m}^2)
+\frac{(4x^2-6x+1)m^2}{(1-x)^2}\delta(s-\tilde{m}^2)
\\&&-(4x^2-2x)\theta(s-\tilde{m}^2)\Big\}
\label{last_eqn}
\non
&=&-\frac{\JJb}{864\pi^2}\frac{s^3-6m^2s^2+6m^4s-40m^6}{(s-4m^2)^2s}\sqrt{1-4m^2/s}, \label{SP2++}
\non
\mathcal{L}_0(M_B^2,s_0)&=&\int_{4m^2}^{s_0} \left[\rho^{pert}(s)+\rho^{\GGa}(s)\right]e^{-s/M_B^2}ds+
\lim_{\eta\to 0^+}\Bigg\{
\non
&&\int_{4m^2(1+\eta)}^{s_0}\left[\rho^{\GGGa}(s)+\rho^{\JJa}(s)\right]e^{-\frac{s}{M_B^2}}ds
-\frac{4m^2}{\sqrt{\eta}}\left[\frac{\GGGb}{768\pi^2}+\frac{\JJb}{576\pi^2}\Bigg]e^{-\frac{4m^2}{M_B^2}}
\right\},
\end{eqnarray}
}
\end{itemize}
in which $m$ is the heavy quark mass, $\delta'(s-\tilde m^2)=\frac{\partial \delta(s-\tilde m^2)}{\partial s}$, $\tilde m^2=\frac{m^2}{x(1-x)}$ where $x$ is a Feynman parameter.

\end{document}